\documentclass{pasa}%

\usepackage{hyperref}
\usepackage{breakurl}
\usepackage{framed}

\usepackage{float}
\usepackage{caption}
\usepackage[subrefformat=parens,labelformat=parens]{subcaption}

\let\stdcaption\caption
\usepackage{float}
\let\caption\stdcaption

\usepackage[authoryear]{natbib}

\usepackage{supertabular}

\usepackage{bm}

\usepackage{color}
\usepackage{color,hyperref}
\definecolor{darkblue}{rgb}{0.0,0.0,0.5}
\definecolor{yell}{rgb}{.8,.9,.0}
\hypersetup{colorlinks,breaklinks,
            linkcolor=darkblue,urlcolor=darkblue,
            anchorcolor=darkblue,citecolor=darkblue}

\usepackage[figuresleft]{rotating}
\usepackage{animate}

\title[MWA ionosphere calibration]{Ionospheric Modelling using GPS to Calibrate the MWA. II: Regional ionospheric modelling using GPS and GLONASS to estimate ionospheric gradients}
\def\Curtin{$^{1}$}
\def\USydney{$^{2}$}
\def\CASS{$^{3}$}
\def\CAASTRO{$^{4}$}
\def\RRI{$^{5}$}
\def\UWA{$^{6}$}
\def\Victoria{$^{7}$}
\def\ANU{$^{8}$}
\def\UMelbourne{$^{9}$}

\author[Arora et al.]{B.~S.~Arora\Curtin, 
J.~Morgan\Curtin, 
S.~M.~Ord\Curtin,
S.~J.~Tingay\Curtin,
M.~Bell\USydney,
J.~R.~Callingham\CASS$^,$\USydney$^,$\CAASTRO, 
K.~S.~Dwarakanath\RRI,
B.-Q.~For\UWA,
P.~Hancock\Curtin$^,$\CAASTRO, 
L.~Hindson\Victoria,
N.~Hurley-Walker\Curtin,
M.~Johnston-Hollitt\Victoria, 
A.~D.~Kapi\'{n}ska\UWA$^,$\CAASTRO,
E.~Lenc\USydney$^,$\CAASTRO, 
B.~McKinley\ANU$^,$\CAASTRO, 
A.~R.~Offringa\ANU,
P.~Procopio\UMelbourne$^,$\CAASTRO, 
L.~Staveley-Smith\UWA$^,$\CAASTRO, 
R.~B.~Wayth\Curtin$^,$\CAASTRO, 
C.~Wu\UWA
and Q.~Zheng\Victoria \\
\\
\affil{$^{1}$International Centre for Radio Astronomy Research, Curtin University, Bentley, WA 6102, Australia}%
\affil{$^{2}$Sydney Institute for Astronomy, The University of Sydney, 44 Rosehill Street, Redfern, NSW 2016, Australia}%
\affil{$^{3}$CSIRO Astronomy and Space Science (CASS), PO Box 76, Epping, NSW 1710, Australia}%
\affil{$^{4}$ARC Centre of Excellence for All-Sky Astrophysics (CAASTRO), Redfern, NSW 2016, Australia}%
\affil{$^{5}$Raman Research Institute, Bangalore 560080, India}%
\affil{$^{6}$International Centre for Radio Astronomy Research (ICRAR), University of Western Australia, Crawley, WA 6009, Australia}
\affil{$^{7}$School of Chemical \& Physical Sciences, Victoria University of Wellington, Wellington 6140, New Zealand}%
\affil{$^{8}$Research School of Astronomy and Astrophysics, Australian National University, Canberra, ACT 2611, Australia}%
\affil{$^{9}$School of Physics, The University of Melbourne, Parkville, VIC 3010, Australia}%
}
\jid{PASA}
\doi{10.1017/pas.\the\year.xxx}
\jyear{\the\year}

\begin{document}%
\begin{abstract}
We estimate spatial gradients in the ionosphere using the Global Positioning System (GPS) and GLONASS (Russian global navigation system) observations, utilising data from multiple GPS stations in the vicinity of Murchison Radio-astronomy Observatory (MRO). In previous work the ionosphere was characterised using a single-station to model the ionosphere as a single layer of fixed height and this was compared with ionospheric data derived from radio astronomy observations obtained from the Murchison Widefield Array (MWA). Having made improvements to our data quality (via cycle slip detection and repair) and incorporating data from the GLONASS system, we now present a multi-station approach. These two developments significantly improve our modelling of the ionosphere. We also explore the effects of a variable-height model. We conclude that modelling the small-scale features in the ionosphere that have been observed with the MWA will require a much denser network of Global Navigation Satellite System (GNSS) stations than is currently available at the MRO.

\end{abstract}
\maketitle%
\section{INTRODUCTION }
\label{sec:intro}

The Earth's ionosphere has a significant effect upon radio astronomy observations, in particular at low radio frequencies. Below a critical frequency the ionosphere is opaque \citep{Raw93, Wil14} and can radiate \citep{Dav90}.  Refractive effects manifest themselves as apparent position shifts of celestial radio sources \citep{Wil14}. The ionosphere is dispersive \citep{Dav90} and causes Faraday Rotation of radio waves \citep{Wil14}.  The ionosphere can also cause diffractive effects \citep{Dav90}.\\

The ionosphere is of great interest as a target for research for many reasons and has been the subject of detailed study for decades.  See \citet{Dav90} and \citet{Raw93} for general reviews of the ionosphere and \citet{Tho07} and \citet{Wil14}, for example, for the connection between ionospheric research and radio astronomy.\\

With a new generation of wide-field low radio frequency telescopes now in operation, including the Murchison Widefield Array (MWA) \citep{2013PASA...30....7T}, Low-Frequency Array (LOFAR) \citep{van13}, Precision Array for Probing the Epoch of Reionization (PAPER) \citep{Par10}, and the Long Wavelength Array (LWA) \citep{Ell09}, interest in the effect of the ionosphere in radio astronomy is greatly renewed. This is for two reasons: firstly the new generation of radio telescopes have the ability to probe the ionosphere in unprecedented detail \citep{Loi2015a}. Secondly, because as the new generation of radio telescopes are designed and built, calibration of the effects of the ionosphere become more challenging and characterising its effects radio astronomy observations is critical.\\

The MWA, the low frequency precursor for the SKA located in Western Australia, has 128 aperture array elements (called tiles), has the maximum baseline length of $\sim$3 km, and an extreme wide field-of-view (FoV) capability (25$^{\circ}$ full width at half maximum at 150 MHz) \citep{2013PASA...30....7T}.  These characteristics place the MWA in a regime where different paths through the ionosphere are observed for different sources across the FoV, but the effect of the ionosphere on each array element can be assumed to be the same \citep{Lon09}. However future instruments \citep[e.g. the low frequency component of the Square Kilometre Array (SKA)][]{Hal05} will have much longer baselines, necessitating not only a direction-dependent calibration, but also a different solution for each interferometer element, a far more difficult and computational intensive problem to solve.\\

This motivates us to look at Global Satellite Navigation Systems (GNSS) as a possible source of information on the ionosphere, not only as a direct source of information for calibration (perhaps a low-resolution model that can reduce the parameter space to be searched), but also both for climatology (understanding the range of ionospheric conditions in a statistical sense), and for identifying whether conditions prevailing during a particular observations were favourable for radio astronomy (without taking the much more route of determining this from the radio telescope data itself). In previous work \citep{Aro15} we undertook an initial study of refractive effects due to the ionosphere, as observed by the MWA, and compared them with independent measurements using the Global Positioning System (GPS). Bulk ionospheric gradients causing the refractive effects observed with the MWA were found to agree well with those estimated from GPS observables. The results presented in \citet{Aro15} establish a methodology and show that ionospheric information can plausibly be obtained from GPS, to help calibrate the MWA. \\


The research presented here aims to build on our earlier work.  Before, ionospheric modelling was performed by using data from a single GPS station for any given ionospheric solution. To capture the ionospheric behaviour on finer spatial scales, additional data is required. To this end, we incorporate the GLONASS satellite system into our analysis.  GLONASS currently has 24 active satellites in orbit. Further, we now upgrade our ``single-station'' analysis to a ``multi-station'' analysis, whereby each ionospheric solution is calculated using data from multiple receiving stations.\\

Finally, we explore the effectiveness of various relaxations of the single layer model for ionospheric modelling.  Methods to include spatial and temporal variations into the height of the single layer model are discussed.\\


This paper is organised as follows; Section \ref{sec:prepro} presents the GNSS data pre-processing methodology. In Section \ref{sec:method} the GLONASS system overview and a combined GPS and GLONASS observation model are presented. Further, the effect of the single layer model height on the estimated ionosphere coefficients and methods to incorporate the variation in single layer model height are presented. The multi-station approach to estimate ionosphere coefficients using GPS and GLONASS is presented in Section \ref{sec:method}. Section \ref{sec:mwaiono} presents the summary of obtaining ionosphere gradients from MWA observations as a function of position shifts. The results from the multi-station approach are presented and discussed in Section \ref{sec:results}. The paper is concluded in Section \ref{sec:conc}, where we discuss future directions for this work.

\section{GNSS DATA PRE PROCESSING}
\label{sec:prepro}

In GNSS data pre-processing, discontinuities in the phase and code observables are identified and repaired \citep[][among others]{Lic90,Rem95}. The uncertainties in the GNSS observables are phase cycle-slips and jumps, and multi-path effects. In our previous work \citet{Aro15}, pre-processing of the GNSS observables was applied, however it is discussed in detail here for the first time.
 
\subsection{Cycle-slip detection and repair}

When a receiver tracks a satellite, the integer and fractional number of cycles (total number of wavelengths at the GPS frequency) between the receiver and the satellite are recorded as phase observables. However, the initial number of phase cycles, upon first acquisition of the satellite signal, remain ambiguous. The ambiguities present in the phase observables remain constant for a complete satellite pass, unless a cycle-slip occurs. Cycle-slip can occur for a number of reasons including, temporary blocking of the GNSS signal by a physical obstruction, multi-path effects, high ionospheric activity, and low signal-to-noise ratio \citep{Hof08}. It is important to account for cycle-slips in order to ensure the continuity of carrier phase data on which high precision GNSS applications are dependent.\\

There are three stages for pre-processing of cycle-slips. Firstly, the cycle-slip is detected, secondly its magnitude is quantified, and thirdly it is flagged or accounted for in the observables. The generic approach to cycle-slip detection is by forming linear combinations of observables. During pre-processing using the BERNESE software \citep{Dac07}, a combination of the phase and code observables is formed, also known as Melbourne-W\"{u}bbena combination \citep{Mel85,Wub85}, which allows detection of cycle-slips. However, the noise of the observable in Melbourne-W\"{u}bbena combination is driven by the noise of the code observable, the code observables are found to have a precision of 25 cm or worse. Other combinations, namely, the ``wide-lane'' \citep{Hof08} and ``ionosphere-free'' \citep{Hof08} combinations, although driven by the very precise phase observables, have noise of 5.7 and 3.0 times the original observables, respectively \citep{Dac07}. \\


%
%

The Geometry-Free combination can also be used to detect cycle-slips \citep{Vac15}, the Geometry-Free phase observable, used to detect cycle-slips, is denoted as L4. The noise of the L4 observable is $\sqrt{2}$ times the precision of the phase observables, lower than all of the earlier mentioned linear combinations. In our approach, cycle-slips are detected by using Geometry-Free combination of observables. \\

For a Geometry-Free combination, the new phase observables (L4) has the constant instrumental term, which has ambiguities and other biases, and the ionospheric error. The time difference of the L4 observables, L4$(t)$-L4$(t-1)$ can be used to eliminate all other terms except the variable part of the ionospheric error. Any unusual variation in the ionosphere can be easily flagged for a cycle-slip. Following \citet{Vac15}, the expression for cycle-slip detection, using Geometry-Free phase observables, is given as follows:

\begin{equation}\label{eq:cysl}
|\textnormal{L4}(t) - \textnormal{L4}(t-1)| > k \cdot \sigma_{\textnormal{L4}} + \Delta I_{max}
\end{equation}

where L4 is the Geometry-Free observable formed from GNSS phase observables L1 and L2, $t$ is the observation time, $k$ is a scaling factor,  $\sigma_{\textnormal{L4}}$ is the precision of the $\textnormal{L4}$ observable, and $I_{max}$ is the maximal ionospheric delay. Following \citet{Vac15}, $\Delta I_{max}$ is chosen to be 0.4 $m/hour$ and the factor $k$ as 4. \\

Once the cycle-slip is detected, the hypothesis can further be tested by using the Geometry-Free code observable (P4) for a sufficient number of epochs \citep{Teu98}.\\

The cycle-slip can be repaired by estimating the time propagation of the ionosphere from L4 observables, using the information before and after the slip. The L4 observables being precise, are capable of sensing ionospheric variations as small as $\sim$0.04 Total Electron Content Unit (TECU) (at GPS frequencies, 1 TECU = 10$^{16}$ electrons m$^{-2}$). Considering the location of GNSS receivers we are using (far below the equatorial anomaly), extreme ionospheric variations are not expected. However, ionospheric variations of the order of 1 TECU every 30 seconds can be easily accounted for with this algorithm. \\

The above algorithm, however, is not capable of differentiating whether the slip occurred on L1 or L2 phase observable. Since our software makes use of L4 observables for estimating the ionosphere and other unknowns, estimating cycle-slips at individual frequencies is not a necessary.

\section{IONOSPHERIC MODELLING USING GPS AND GLONASS}
\label{sec:method}

\subsection{GLONASS system overview}
The GLONASS system, currently has 24 operational satellites in its constellation, which continuously transmit dual frequency data centered at frequencies L1$_{0}$ = 1602.0 MHz and L2$_{0}$ = 1246.0 MHz. Each GLONASS satellite transmits on a different frequency using a 15-channel Frequency Division Multiple Access (FDMA) technique. The frequency for each channel is given by L1$_{n} = $L1$_{0} + n \times (9/16) $ MHz and L2$_{n}$ = L2$_{0} + n \times (7/16) $ MHz, where $n$ is the GLONASS channel number, $n=-7,\cdots,0,\cdots,6$. The GLONASS channel number for each satellite can be obtained from the GLONASS broadcast ephemerides file.\\

The GPS and GLONASS systems transmit in different reference times, which needs to be compensated in the code observables, while realising a common time system for processing the data.  However, if the Geometry-Free combination is used, as in our work, the code observable correction is compensated and not required.

\subsection{GPS and GLONASS observation model}
\label{sec:model}

\begin{figure*}
 \centering
 \subcaptionbox{STEC for MRO1 - GPS only\label{fig:STECMRO1062}}{\includegraphics[scale=0.45]{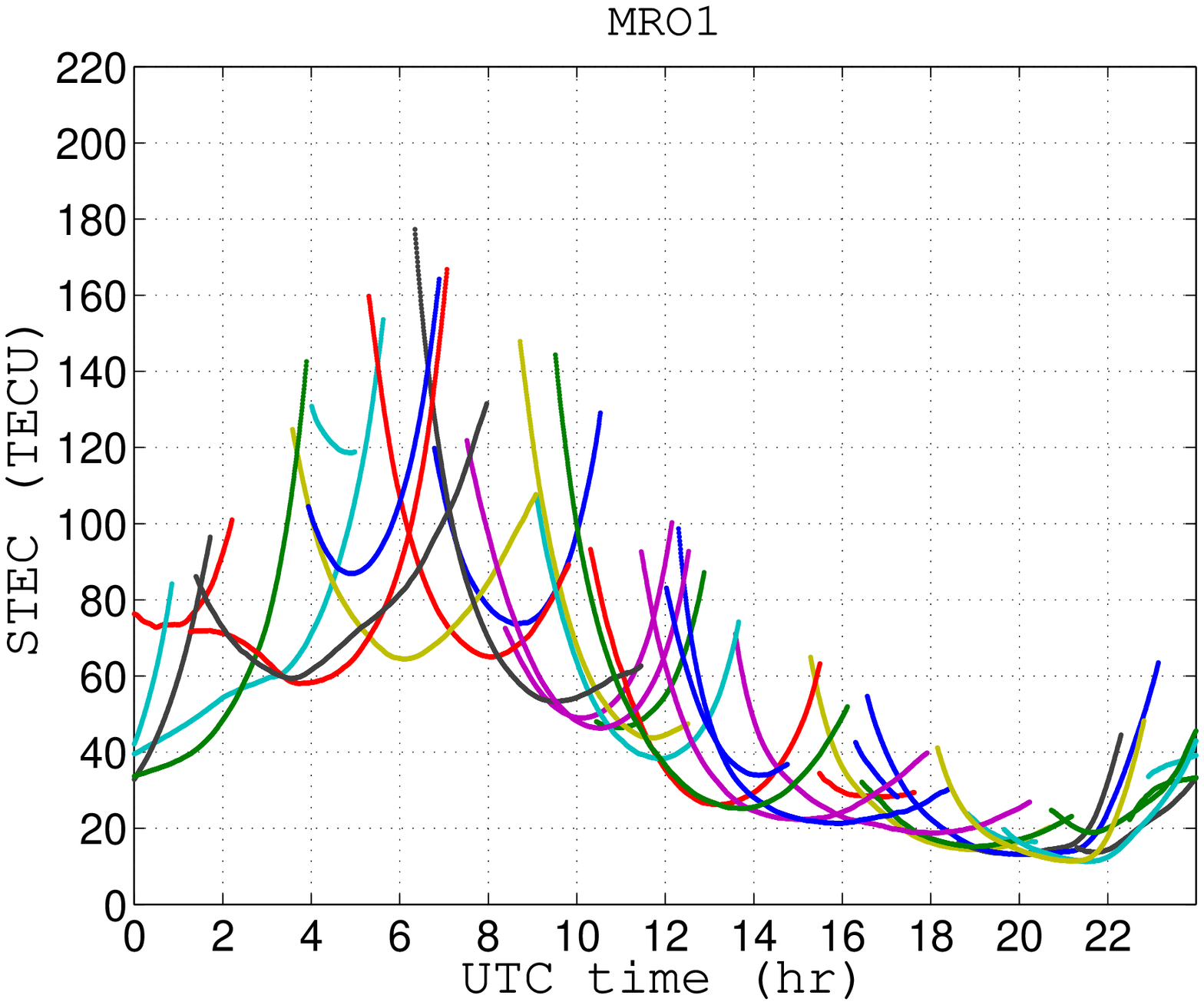}}
 \subcaptionbox{STEC for MEDO - GPS+GLONASS \label{fig:STECMEDO062}}{\includegraphics[scale=0.45]{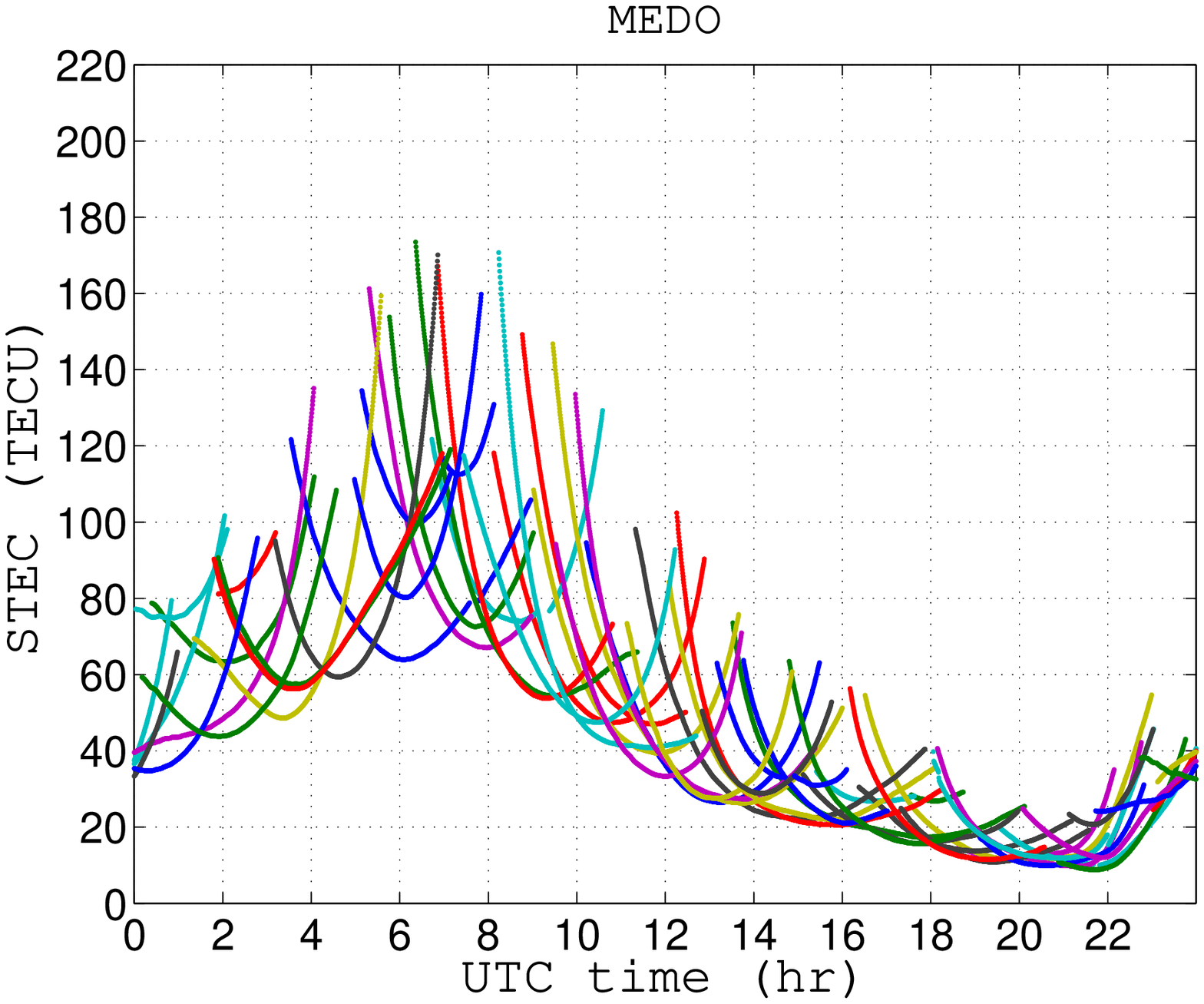}}
     \caption{Retrieved STEC for the MRO1 and MEDO Geoscience Australia (GA) GNSS stations on DOY 062, year 2014.
        \label{fig:retSTEC}}
\end{figure*}

The Geometry-Free GPS observation model is discussed in detail in \citet{Aro15}. We recall it below and append the GLONASS observation model as follows

\begin{equation} \label{eq:sdphase1}
\displaystyle E(\Phi_{r,21}^{Gs}) = \displaystyle \Phi_{r,1}^{Gs}-\Phi_{r,2}^{Gs} =  -  \iota_{r,21}^{Gs} + \mathrm{C}_{r}^{Gs} 
\end{equation}
\begin{equation}  \label{eq:sdphase2}
\displaystyle E(P_{r,21}^{Gs}) = \displaystyle P_{r,1}^{Gs}-P_{r,2}^{Gs} =  \iota_{r,21}^{Gs} + c \cdot(d_{r,21} - d^{Gs}_{,21})
\end{equation} 
\begin{equation} \label{eq:sdphase11}
\displaystyle E(\Phi_{r,21}^{Rs}) = \displaystyle \Phi_{r,1}^{Rs}-\Phi_{r,2}^{Rs} =  -  \frac{\mu_{21}^{R}}{\mu_{21}} \iota_{r}^{s} + \mathrm{C}_{r}^{Rs}
\end{equation}
\begin{equation}   \label{eq:sdphase22}
\displaystyle E(P_{r,21}^{Rs}) = \displaystyle P_{r,1}^{Rs}-P_{r,2}^{Rs} =  \frac{\mu_{21}^{R}}{\mu_{21}} \iota_{r}^{s} + c \cdot d^{Rs}_{r,21}
\end{equation}

where $E(\cdot)$ is the expectation operator, $\Phi$ is the phase observable, $P$ is the code observable, subscript ${}_{r}$ indicates receiver, ${}_{1}$, ${}_{2}$ and ${}_{21}$ indicates GNSS frequency/frequency combinations corresponding to phase (or code) observables, L1 (or C1), L2 (or P2) and L4 (or P4), respectively. Superscripts ${}^{Gs}$, ${}^{Rs}$ indicate GPS and GLONASS satellites, respectively, $c \cdot(d_{r,21})$ and $c (d^{Gs}_{,21})$ are the GPS receiver and satellite Differential Code Biases (DCBs), respectively, $\mu_{21}$ is the GPS frequency coefficient given as, $\mu_{21}= \mu_{1} - \mu_{2}$ and $\mu_{1} =  \displaystyle \frac{1}{f_{1}^{2}}$, $\mu_{2} =  \displaystyle \frac{1}{f_{2}^{2}}$, ${f_{1}}$ and ${f_{2}}$ are GPS frequencies at L1 and L2. Similarly, the GLONASS frequency coefficient is given by $\mu_{21}^{R}$. \\

The instrumental biases and other unknowns are estimated for each GNSS receiver using the method of least squares with a Kalman filter, as described in \citet{Aro15}. The precision of the time-constant parameters propagate as the inverse of the square-root of the number of epochs ($n$), $\sigma = 1/\sqrt{n}$. For a continuous satellite arc, $n$ is very large, $\sim$100 or more. This results in a very precise estimation of time-constant parameters. \\

The line-of-sight Total Electron Content (TEC) between receiver and the satellite, is also known as Slant TEC ($STEC$). $STEC$ can be retrieved from L4 phase observables ($\Phi_{r,21}$). In our work, we retrieve the $STEC$ for both GPS and GLONASS satellites, by substituting for the time-constant parameters for L4 observables, estimated using the single-station approach. Figure \ref{fig:retSTEC} presents the retrieved $STEC$ for MRO1 and MEDO Geoscience Australia (GA) GNSS stations on DOY 062, refer Figure \ref{fig:MWAIPP2014062} and Table \ref{tab:descGNSSnet} for details of all the stations. \\

\subsection{Single-station versus multi-station approach}

\begin{figure*}
 \centering
 \subcaptionbox{$VTEC$ as a function of $H_{ion}$\label{fig:vtecvshion}}
{\includegraphics[scale=0.45]{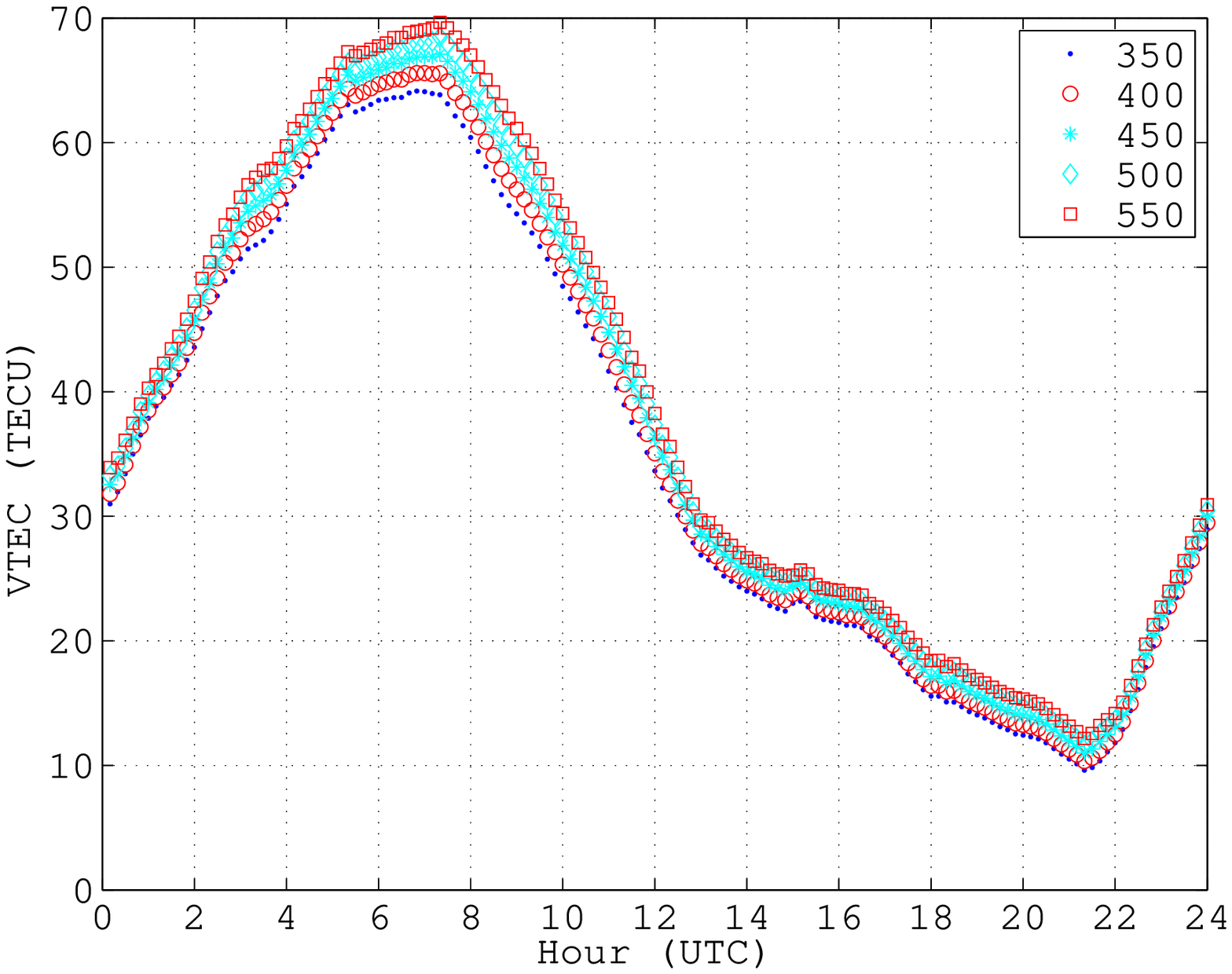}}
\subcaptionbox{Receiver DCB as a function of $H_{ion}$\label{fig:rdcbvshion}}
{\includegraphics[scale=0.45]{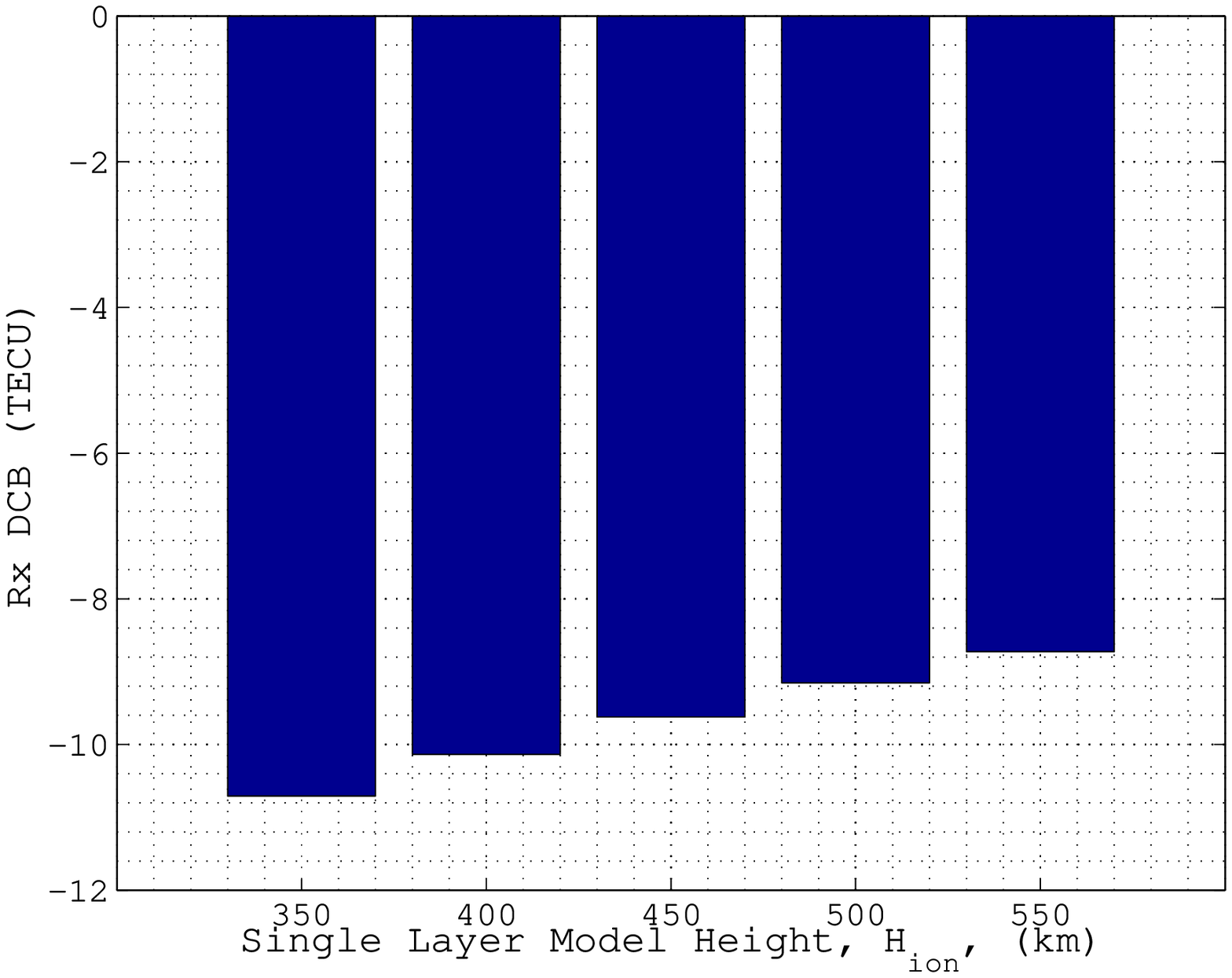}}
     \caption{Effect on estimated $VTEC$ and receiver DCBs by the choice of $H_{ion}$, $H_{ion}$ is varied between 350 to 550 km in steps of 50 km. Note the average precision of $VTEC$ is $\sim$0.03 TECU. (a) $VTEC$ as a function of $H_{ion}$. (b) Receiver DCB as a function of $H_{ion}$.
        \label{fig:hioneffect}}
\end{figure*}

In a single-station approach, the ionospheric coefficients and time constant parameters, namely the frequency dependent receiver and satellite biases on code observable, also known as DCBs, and the constant phase term (containing the ambiguities and other biases) are estimated using the observables from a single-station. Due to the limited number of observations, the ionospheric coefficients are estimated at an interval of 10 minutes to allow sufficient robustness. The ionosphere gradients show artificially high levels of variation as a satellite sets and rises, again due to the limited number of observations used.\\

For multi-station ionosphere modelling, the GNSS model can be designed such that all the parameters (ionosphere and other time-constant biases) are estimated in a multi-station mode. This approach adds constraints to the time-constant satellite-specific bias parameters, namely, the GPS satellite DCBs. However, the multi-station approach must account for different number of satellites being visible, for different receivers at any given time. Another feasible approach is to consider only satellites that are visible to all receivers at a given time, however this can result in loss of information. In this work, the multi-station modelling is performed using the retrieved $STEC$ for each GNSS receiver, which eliminates the need to estimate any time constant parameter.\\

In our work, the retrieved $STEC$ is derived from the precise phase observables only, in contrast to the generic approach of using phase-smoothed-code observables \citep{Gao02,Che13}. In the phase-smoothed-code approach, the phase as well as the code observables are used to retrieve the $STEC$, by substituting for the receiver and satellite DCBs. Also that, the noise of retrieved $STEC$ from the phase-smoothed-code approach is driven by the noise of the code observable. By using the phase observables to retrieve $STEC$, the noise of the $STEC$ is driven by the noise of the phase observable. In our approach, the time constant biases in the phase observable, specific to each receiver and satellite are estimated during the single-station approach, with sufficient precision that complements the noise of the phase observable. These constant phase biases are then subtracted to retrieve the $STEC$ for each receiver-satellite pair. By using retrieved $STEC$, the ionospheric coefficients can be estimated at a higher temporal resolution, in our work, the ionospheric coefficients are estimated every 2 minutes.\\


\subsection{Effective height of the ionospheric layer}

For ionospheric modelling using a single layer model, the ionospheric electron density is assumed to be concentrated at a fixed height, $H_{ion}$. $H_{ion}$, is used to compute the obliquity factor (mapping function) and the coordinates of the Ionospheric Pierce Point (IPP). \\

\begin{figure*}[htbp]
 \centering
 \subcaptionbox{EW gradient v/s HION \label{fig:EWvsHION062}}{\includegraphics[scale=0.45]{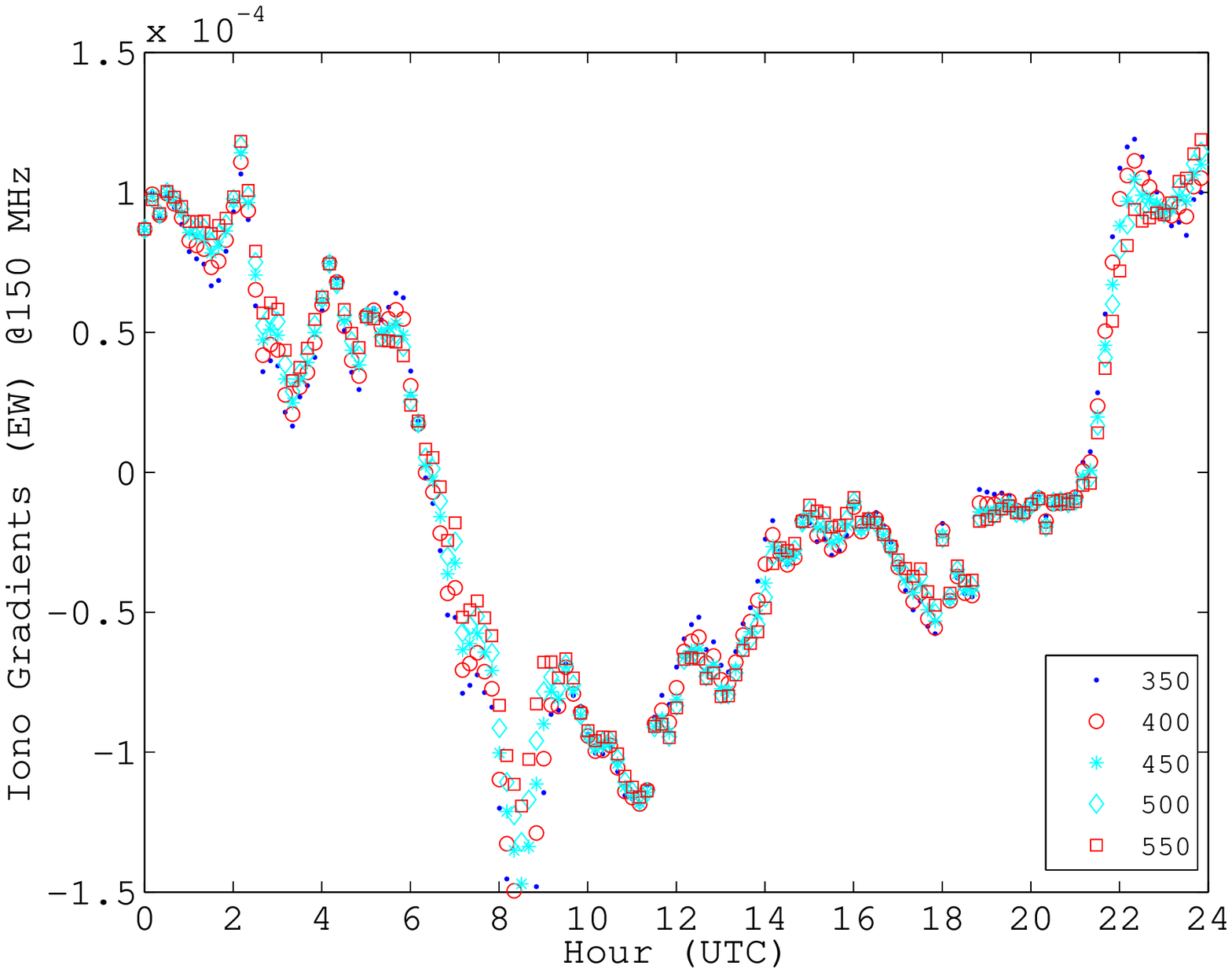}}
\subcaptionbox{NS gradient v/s HION\label{fig:NSvsHION062}}{\includegraphics[scale=0.45]{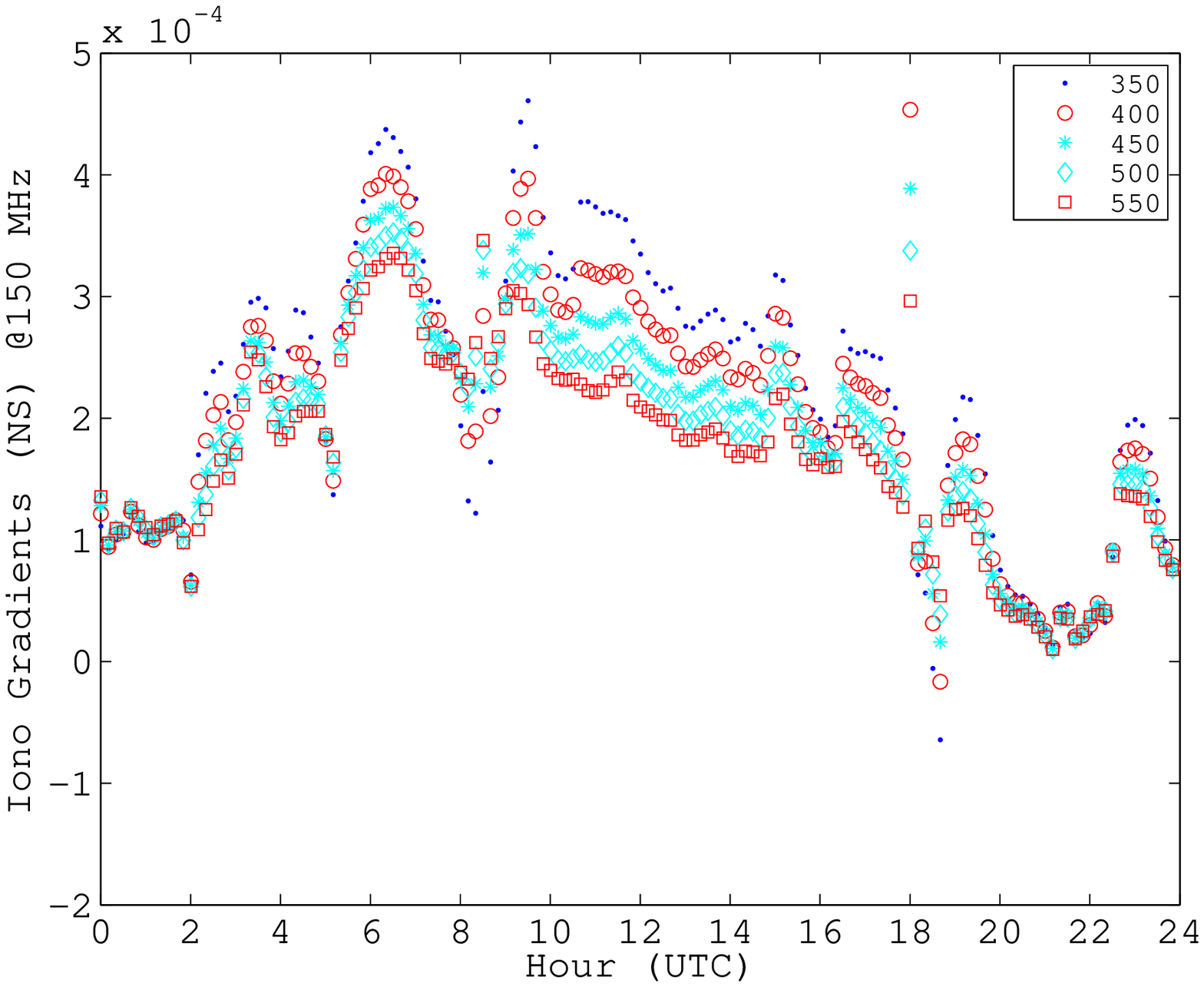}}
     \caption{Effect on estimated ionosphere gradients by the choice of $H_{ion}$, $H_{ion}$ is varied between 350 to 550 km in steps of 50 km.
        \label{fig:hioneffectongrad}}
\end{figure*}

\begin{figure*}
 \centering
{\includegraphics[scale=0.7]{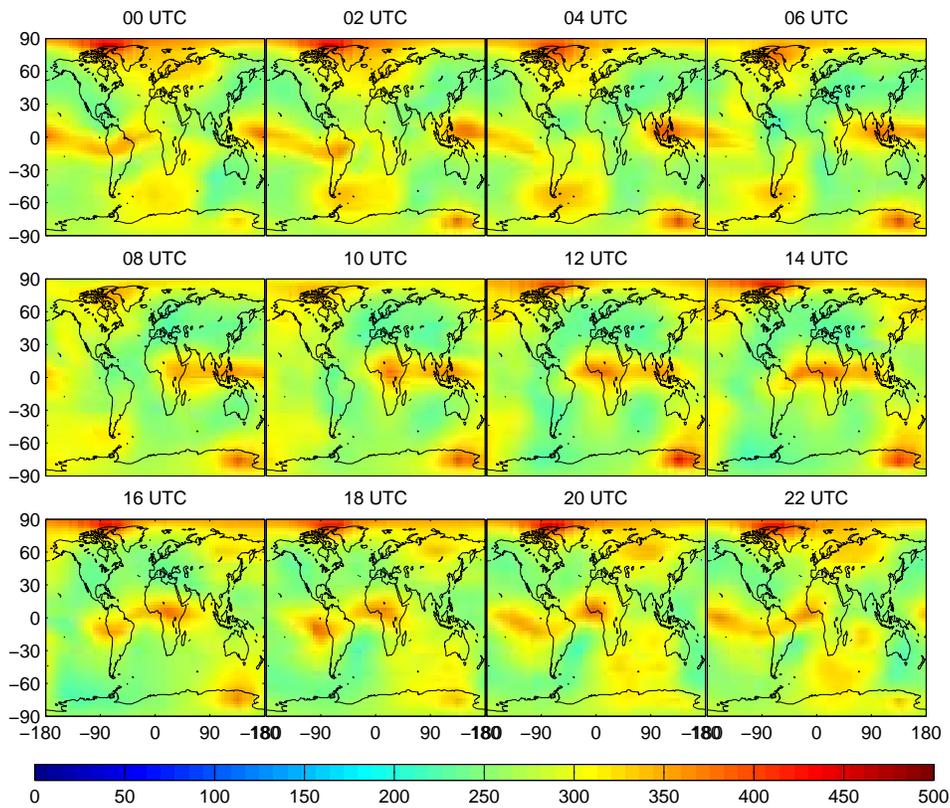}}
     \caption{HmF2 global map from IRI-Plas model for DOY 062, 2014.
        \label{fig:hmf2map}}
\end{figure*} 

To understand the effect of $H_{ion}$ on estimated ionospheric coefficients and receiver DCBs, $H_{ion}$ was varied between 350 and 550 kms in steps of 50 km. GPS observables for GA station MRO1 for DOY 062, year 2014 were used for this analysis. GPS data were processed using a single-station, single-layer ionospheric modelling approach. Figure \ref{fig:hioneffect}\subref{fig:vtecvshion} presents the estimated values of $VTEC$ (Vertical TEC) for different values of $H_{ion}$. The differences in $VTEC$ lie between $\sim$0.5 and $\sim$1 TECU at different times during the day. This effect is absorbed by the receiver DCBs, the receiver DCBs are affected by an amount corresponding to the maximum constant difference in $VTEC$ over 24 hours, that is $\sim$0.5 TECU  (Figure \ref{fig:hioneffect}\subref{fig:rdcbvshion}). Hence, selection of the value of $H_{ion}$ plays a significant role in ionosphere modelling and has an effect on the estimated receiver DCBs. The ionosphere gradients, however, do not seem to be significantly affected when the height is varied by a constant value, refer Figure \ref{fig:hioneffectongrad}. It is important to incorporate the spatial variations, if they are significant, to the ionospheric single layer height in order to observe any significant change in the gradients.\\

A more realistic representation of the single layer height would be to account for the effective height of the ionospheric layer, $H_{eff}$. $H_{eff}$ is the height at which the electron density reaches its median value, and is a function of location and time. $H_{eff}$ can be deduced from the ionospheric profiles. The ionospheric profiles can be obtained from empirical models like the International Reference Ionosphere (IRI) \citep{Bil14}. However, for IRI model, the maximum height for the ionospheric profiles are limited to 2000 km. Ionospheric-Plasmaspheric models like the Parameterised Ionospheric Model (PIM) \citep{Dan95} and extension of IRI to plasmasphere (IRI-Plas) \citep{Gul12,Gul13} model the ionosphere up to plasmaspheric altitudes (above 20,000 km). In our work, we make use of IRI-Plas model, available as an external software, to generate ionospheric profiles.\\

Though $H_{eff}$ can be deduced from the ionospheric profiles, there is no direct source of information of this parameter, regarding its spatial variation. $H_{eff}$ can also be related to $hmF2$, $hmF2$ is the height at which maximum ionisation is reached, which lies in the F2 region. The spatial variation of $hmF2$ can be obtained from the global $hmF2$ maps generated by IRI-Plas (refer Figure \ref{fig:hmf2map}), available for download\footnote{ftp://ftp.izmiran.rssi.ru/pub/izmiran/SPIM/Maps/hmF2/}. The temporal and spatial resolution of $hmF2$ global maps is 1 hour and 5$^{\circ}$/2.5$^{\circ}$ in longitude and latitude, respectively.\\

\begin{figure}
 \centering
{\includegraphics[scale=0.5]{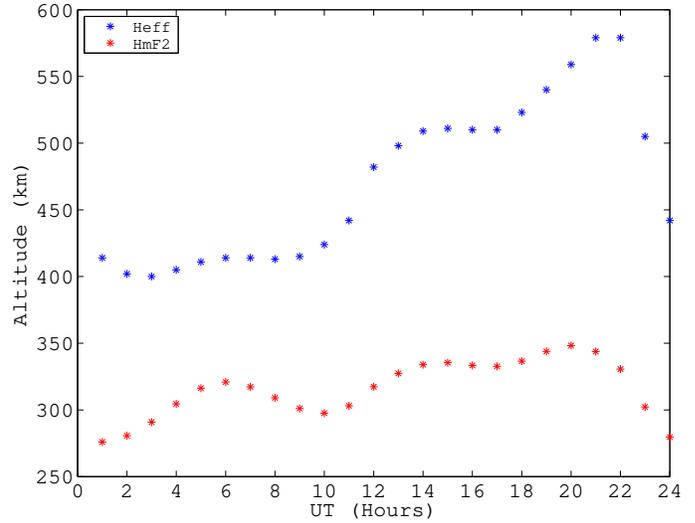}}
     \caption{Temporal variation of $hmF2$ and $H_{eff}$ obtained from IRI-Plas ionosphere profiles at the Taylor series expansion point.
        \label{fig:hmf2heff}}
\end{figure}

The effective height, $H_{eff}$, though is different from $hmF2$, however can be related to $hmF2$. Ionospheric profiles generated using IRI-Plas model show that $H_{eff}$ is greater than $hmF2$, refer Figure \ref{fig:hmf2heff}. It can be noted from Figure \ref{fig:hmf2heff}, the difference of $H_{eff}$ and $hmF2$ increases greatly during the night time, since the electron density in the F2 region decreases and hence $H_{eff}$ increase significantly due to the plasmaspheric electron density. A discussion on $H_{eff}$ and $hmF2$ can be found in \citet{Kom96}. For generating global TEC maps, a variable height of the single layer model has been used for each of the ground stations \citep{Kom98}. In our work, since we make use of regional GNSS network to determine ionospheric gradients, variable height for each receiver-satellite pair is computed.\\

To make use of the effective height, $H_{eff}$, as an input to a single layer model, a method for determining the $H_{eff}$ for all the IPPs is required. For sufficiently small regional scales, as in our work, the offset between $H_{eff}$ and $hmF2$ can be computed at the central location of the network. The only significant variation between $H_{eff}$ and $hmF2$ is due to the solar time, which can be compensated for different IPPs within the network, as an argument of longitude of the IPP. In our work, a constant offset between $H_{eff}$ and $hmF2$ is computed for the Taylor series expansion point. The Taylor series expansion point is the MWA look direction, and remains constant throughout the day (entire observation period). This offset is then applied to the  $hmF2$ values corresponding to satellite IPPs, obtained from the $hmF2$ global maps, refer Figure \ref{fig:hmf2map}.\\

\subsection{GNSS multi-station modelling using retrieved $STEC$}
\label{sec:msmodel}
\begin{figure}
 \centering
{ \includegraphics[scale=0.8]{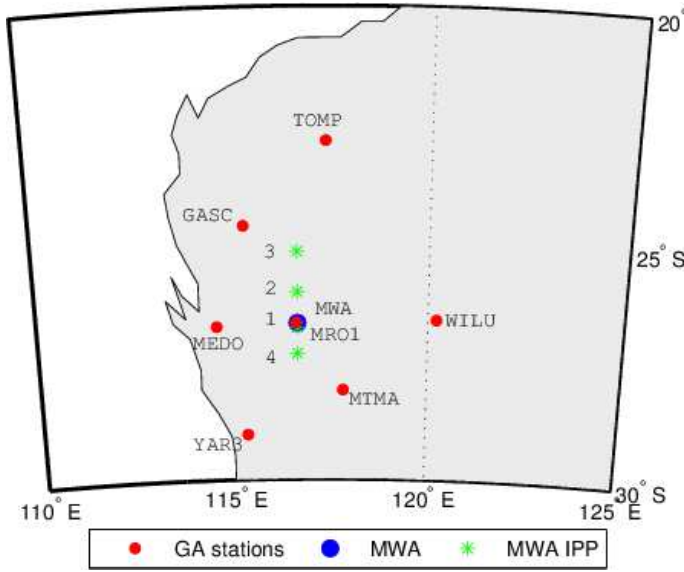}}  
     \caption{Selected GNSS station locations from Geoscience Australia's network (red), MWA location (blue) and MWA IPP (green) for the four MWA observation nights (DOY 062, 063, 065 and 075 marked by 1 to 4, respectively)
        \label{fig:MWAIPP2014062}}
\end{figure}

The GA GNSS stations in the near vicinity of MRO are selected to perform regional modelling. The data from selected GA stations, namely, MRO1, WILU, MTMA, YAR3, MEDO, GASC and TOMP were used for modelling. Figure \ref{fig:MWAIPP2014062} presents the location of the selected GNSS stations.  The details of each GNSS station are given in Table \ref{tab:descGNSSnet}.\\

\begin{table*}
\caption{Description of the selected GA GPS/GNSS stations and the MWA. Data were available for all the four observing sessions, DOY 062, 063, 065, and 075, year 2014. The acronyms given under GNSS, $G$ and $GR$ stand for GPS only and GPS+GLONASS, respectively. } 
\begin{center}
\begin{tabular*}{\textwidth}{@{}c\x c\x c\x c\x c\x c\x c@{}}
\hline \hline
 Station & Receiver type & Antenna type & GNSS & Observables & Location \\
         &               &              &      & used        & (degrees) \\
\hline
 MRO1 & TRIMBLE NETR9 & TRM59800.00 & $G$ & L1, L2, C1, P2 & 26.70$^{\circ}$ S 116.37$^{\circ}$ E \\
 MTMA & LEICA GRX1200+GNSS & LEIAR25.R3 & $GR$ &L1, L2, C1, P2 & 28.11$^{\circ}$ S 117.84$^{\circ}$ E \\
 YAR3$^a$ & LEICA GRX1200GGPRO & LEIAR25   & $GR$ & L1, L2, C1, P2 &29.04$^{\circ}$ S 115.34$^{\circ}$ E \\
 WILU & LEICA GRX1200+GNSS & LEIAR25.R3 & $GR$ & L1, L2, C1, P2 &26.62$^{\circ}$ S 120.21$^{\circ}$ E \\
 MEDO & LEICA GRX1200+GNSS & LEIAR25.R3 & $GR$ & L1, L2, C1, P2 &26.76$^{\circ}$ S 114.61$^{\circ}$ E \\
GASC & LEICA GRX1200+GNSS & LEIAR25.R3 & $GR$ & L1, L2, C1, P2 &24.63$^{\circ}$ S 115.34$^{\circ}$ E \\
TOMP & LEICA GRX1200+GNSS & LEIAR25.R3 & $GR$ & L1, L2, C1, P2 &22.85$^{\circ}$ S 117.40$^{\circ}$ E \\
 MWA & - & - & - & - &26.70$^{\circ}$ S 116.67$^{\circ}$ E  \\
\hline \hline
\end{tabular*}\label{tab:descGNSSnet}
\end{center}
\tabnote{$^a$ Partial data available, from 00:00:00 UTC to 18:07:00 UTC on DOY 062, year 2014}
\end{table*}

\begin{figure}[h]
 \centering
\includegraphics[scale=0.32]{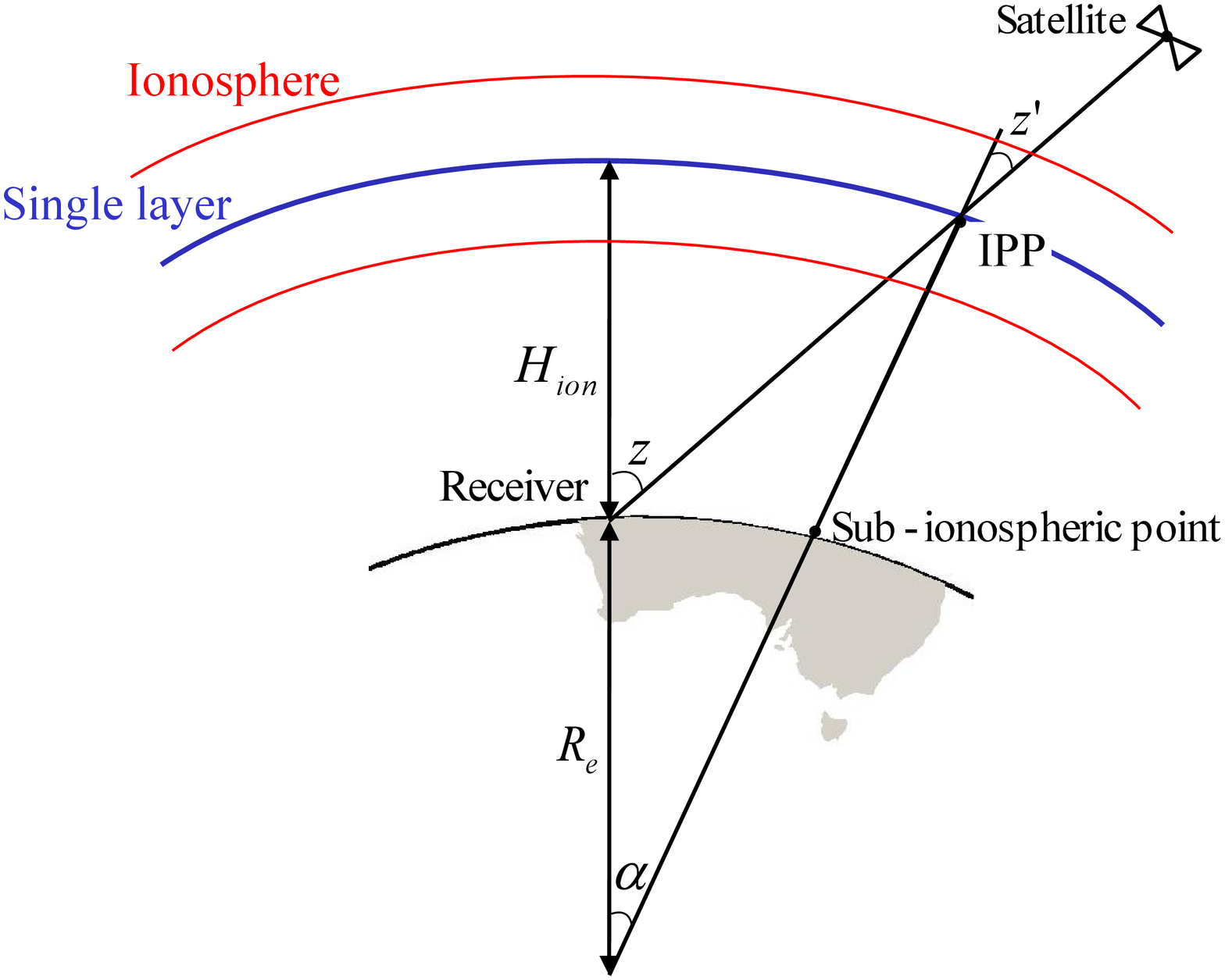}
     \caption{Ionosphere single layer model representation.
        \label{fig:slm}}
\end{figure}

The retrieved $STEC$ for all the GPS and GLONASS satellites are used for regional ionospheric modelling. For a single layer model assumption, $STEC$ can be related to the $VTEC$ using a simple mapping function at a fixed height, refer Figure \ref{fig:slm}. The obliquity factor or the mapping function, $ F^{s}$, is discussed in detail in our earlier paper \citep{Aro15}, the geometry of the obliquity factor is illustrated in Figure \ref{fig:slm}. We briefly summarise the mapping function equation as follows

\begin{eqnarray}\label{eq:Fs}
\left. \begin{array}{lll}
\displaystyle STEC  &=& \displaystyle VTEC \cdot F^{s} \\
\\
\displaystyle F^{s} &=& \displaystyle \frac{1}{\cos(z')} =\frac{1} {\sqrt{1-\sin^{2} z'}} \\
\\
 \displaystyle \sin z' &=& \displaystyle \frac{R_{e}}{R_{e}+H_{ion}} \sin(z)
\end{array}\right\}
\end{eqnarray}

\noindent where $z'$ is the zenith angle at the IPP, $R_{e}$ is the mean radius of the Earth, considered to be 6371 km and assuming a spherical Earth, $H_{ion}$ is the height at the sub-ionospheric  point, and $z$ is the zenith angle of the satellite as seen by the receiver.\\

Two models are considered for the single layer height; first assuming a fixed height of 450 km ($H_{ion}$); second the spatially and temporally varying height inferred from the IRI-Plas model is used ($H_{eff}$).\\

The $VTEC$ is further modelled using a Taylor series polynomial expansion, the expansion point being the MWA IPP. A second order polynomial function is used to model the $VTEC$, given as follows

\begin{eqnarray}\label{eq:polmod2} \nonumber
VTEC (\varphi_{m},s) &=& VTEC_{0} +  (\varphi_{m} - \varphi_{m_{0}})f' \varphi + (s - s_{0})f' s +\\ \nonumber
\\ \nonumber
& & + (\varphi_{m} - \varphi_{m_{0}})^{2}f'' \varphi_{m} \varphi_{m}  + (s - s_{0})^{2}f''  s s \\\nonumber
\\
&& + (\varphi_{m} - \varphi_{m_{0}}) (s - s_{0}) f'' \varphi_{m} s .
\end{eqnarray}

The Sun fixed longitude, $s$, is related to the local solar time ($LT$) as $s = \lambda_{m} + LT - \pi$, where $\lambda_{m}$ is the geomagnetic longitude at IPP, $LT$ is in radians, $VTEC_{0}$ is the $VTEC$ at the central location in the network and $ f' s, \ f' \varphi_{m}, \ f'' s s, \  f'' \varphi_{m} \varphi_{m}, \  f'' \varphi_{m} s $ are the first and second order derivatives of $VTEC$ along the Sun fixed longitude and latitude, respectively.\\

\begin{figure}
 \centering
{\includegraphics[scale=0.7]{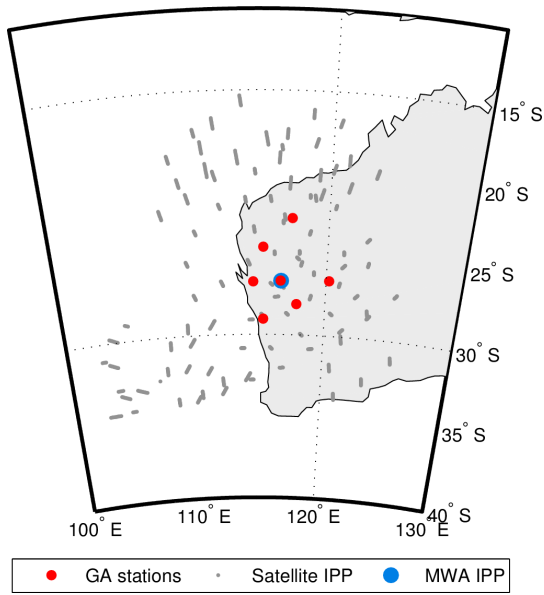}}
     \caption{A snapshot of MWA IPP (blue), GA GNSS stations (red), and satellite IPPs  for 5 minutes (10 epochs) during MWA observations (gray) in Earth-fixed reference frame, on DOY 062, 2014. The MWA IPP is considered for the Taylor series expansion point.
        \label{fig:netipp}}
\end{figure}

Figure \ref{fig:netipp} presents a snapshot of the satellite IPPs for all the GA GNSS stations and MWA IPP locations. 

\section{MWA IONOSPHERE}
\label{sec:mwaiono}
In order to validate our modelled ionosphere against radio astronomy data, precisely the same data were used as in our previous work. Please see \citet{Way15} and \citet{Hur16} for a full description of the GaLactic and Extragalactic All-sky Murchison Widefield Array (GLEAM) survey, and \citet[][Section 4]{Aro15} for a detailed description of how these MWA data may be compared with GPS data. For clarity, the essentials are presented again below.\\

The observations we have used, operate in drift-scan mode, where the telescope remains pointed at a single point on the meridian as radio sources drift past. The telescope also cycles through five frequency bands spanning the range 73--230\,MHz, with each band observed for two minutes in every 10 minutes. Each two-minute observation is then processed to produce one image for each of four neighbouring subbands which make up the full instantaneous bandwidth of 30.72 MHz. Each of these images will typically contain many hundreds of radio sources, most of which are unresolved, and almost all of which are already known from previous surveys.\\

A catalogue of sources has been compiled which are expected to be bright and unresolved with the MWA (Harvey et al. in prep.). For the brightest 100 sources in each image, an elliptical Gaussian is fitted to a small subset of pixels corresponding to the \textit{a priori} location of that source. This gives us the location of the source on the sky at each of four different frequencies.  By fitting the change in location of the source as a function of $\lambda^2$, we can determine the magnitude and direction of ionospheric refraction in a way that is blind to errors in the a prior location of the source, or instrumental or imaging effects. The ionospheric shift is then averaged over all sources within the field of view.\\

The final result is therefore a time-series of ionosphere gradients, in both north-south and east-west directions, averaged over all sources within the MWA field of view. The centre of this field of view is taken to be the MWA pierce point. Only the lowest-frequency band was used, giving a time resolution of 10 minutes.

\section{RESULTS AND DISCUSSION}
\label{sec:results}

\subsection{Comparisons of ionosphere gradients - single-station v/s multi-station approach}

\begin{figure*}[htbp]
 \centering
 \subcaptionbox{EW gradient v/s longitudinal seperation \label{fig:EWvssep062}}{\includegraphics[scale=0.45]{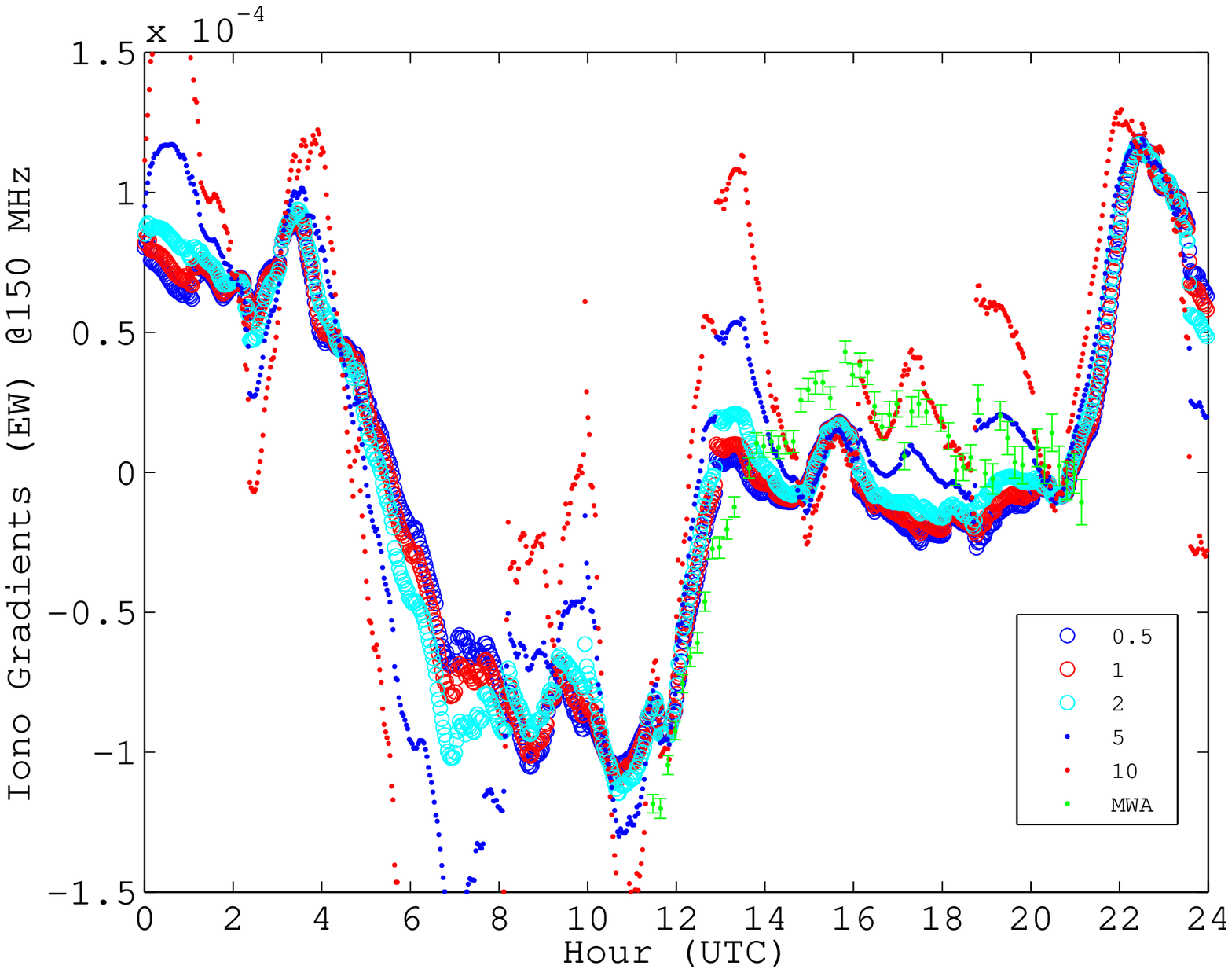}}
\subcaptionbox{NS gradient v/s latitudinal seperation\label{fig:NSvssep062}}{\includegraphics[scale=0.45]{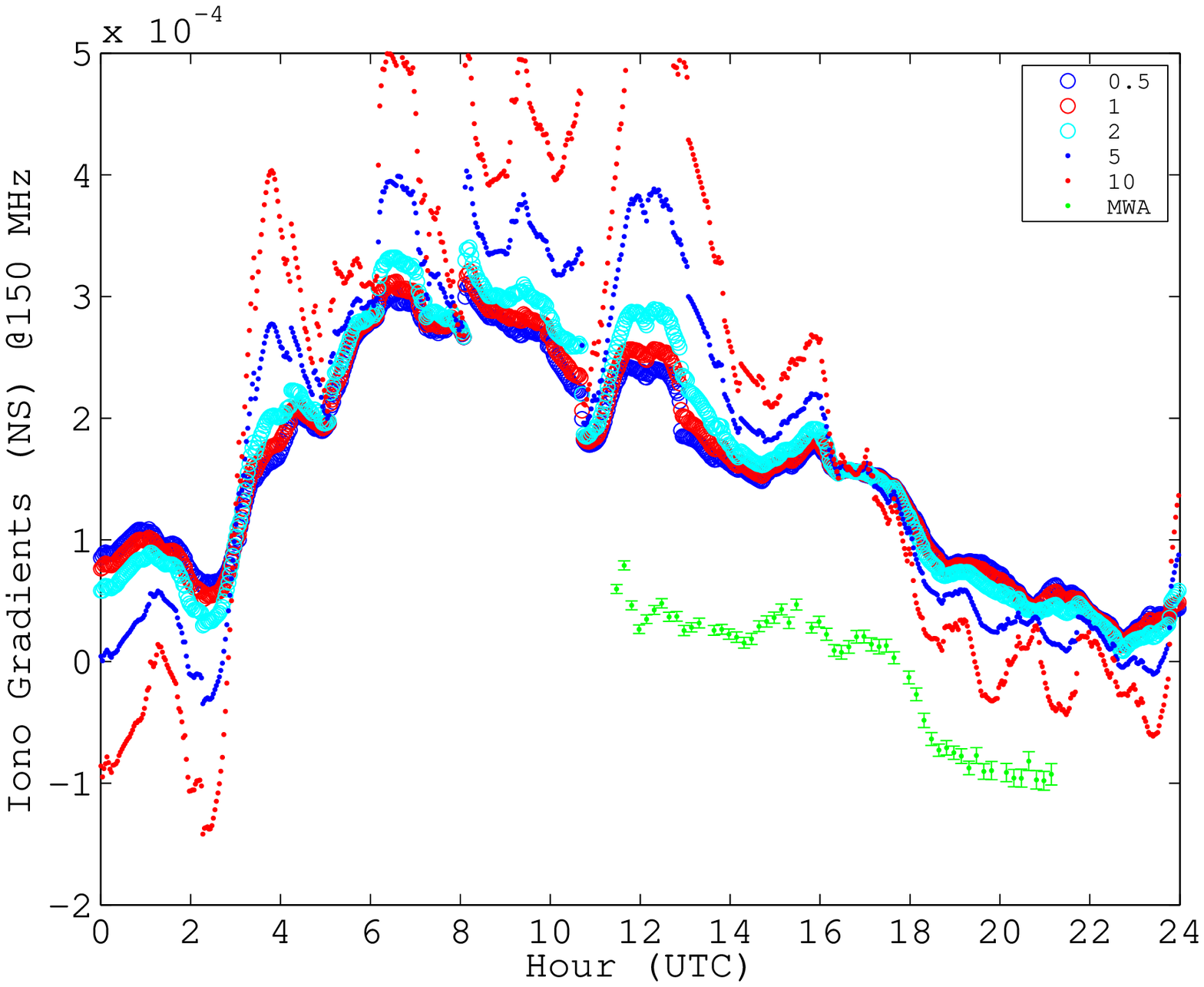}}
     \caption{Effect on estimated ionosphere gradients by the choice of the latitude/longitude separation, the variation for the latitude/longitude was done between 0.5$^{\circ}$ to 10$^{\circ}$. (a) EW gradient v/s longitudinal seperation. (b) NS gradient v/s latitudinal seperation.
        \label{fig:gradsepvar}}
\end{figure*}

The gradients obtained from MWA observations in Right Ascension (East-West or EW direction) and Declination (North-South or NS direction), were computed from the ionospheric coefficients given in equation \eqref{eq:polmod2}. The ionospheric coefficients were estimated by considering the centre of MWA FoV as the expansion point. The gradients can further be calculated by considering a latitude/longitude separation, on scales similar to the MWA FoV, around the expansion point.The MWA would see a FoV of width $\sim$200 km at an altitude of 450 km, and the MWA derived gradients represent the bulk shift over the entire FoV. The ionospheric coefficients obtained using GPS and GLONASS observations were used to compute the EW and NS gradients, over one degree of latitude/longitude (1$^{\circ}$ $\cong$100 km) either side of the expansion point. The variation of latitude/longitude separation was carried for 0.5$^{\circ}$ to 10$^{\circ}$ in order to understand its effect on the computed gradients, refer Figure \ref{fig:gradsepvar}. The gradients computed from GNSS, using the latitude/longitude separation considerations, that closely represent the spatial scales of the MWA FoV, that is 0.5$^{\circ}$ to 2$^{\circ}$, though exhibit some variation, however the effect is modest, refer Figure \ref{fig:gradsepvar}.\\

\begin{figure*}
 \centering
 \subcaptionbox{Single-station - GPS only, DOY 062\label{fig:EWMRO1062}}{\includegraphics[scale=0.33]{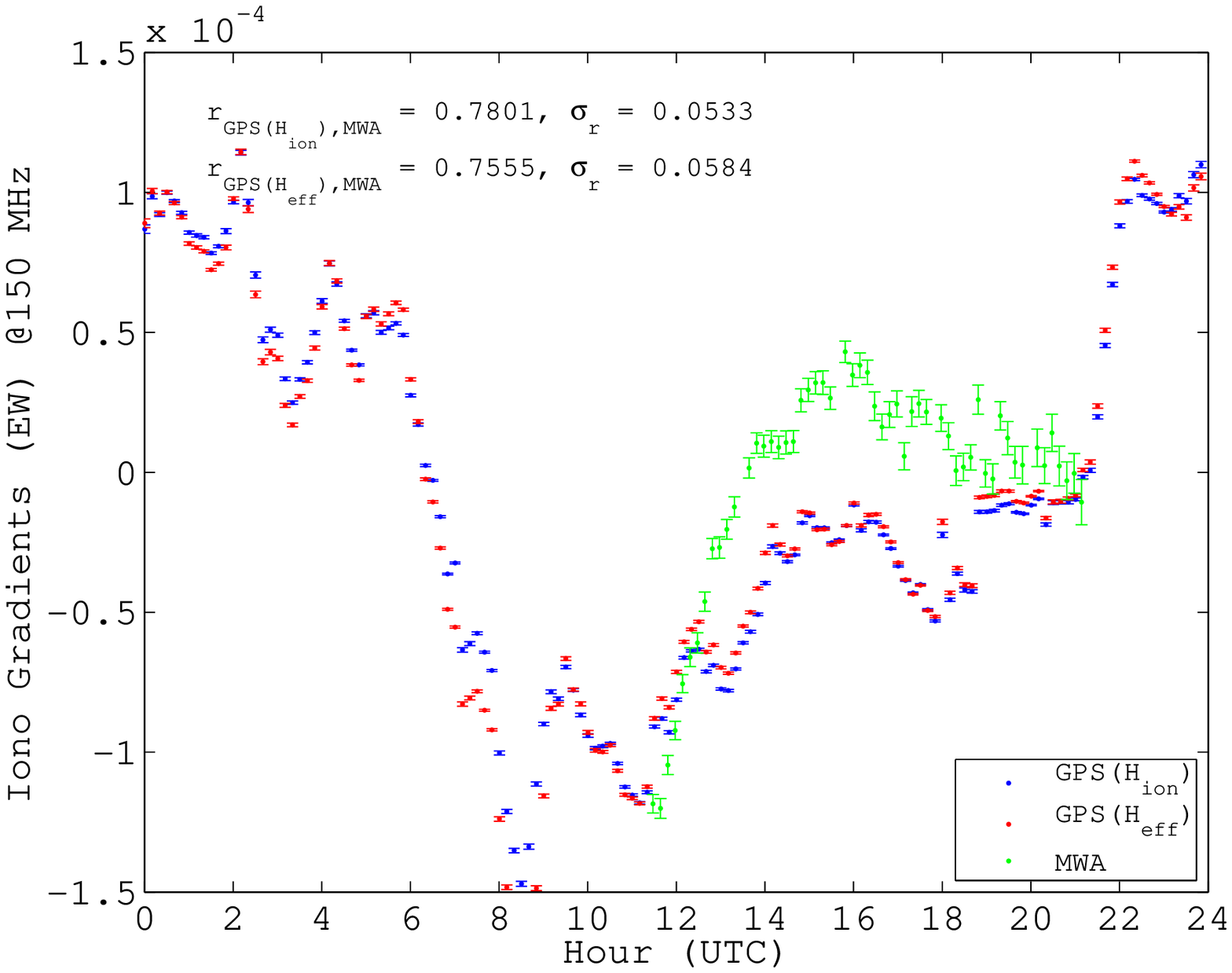}}
\subcaptionbox{Multi-station - GR, DOY 062 \label{fig:ewGR062}}{\includegraphics[scale=0.33]{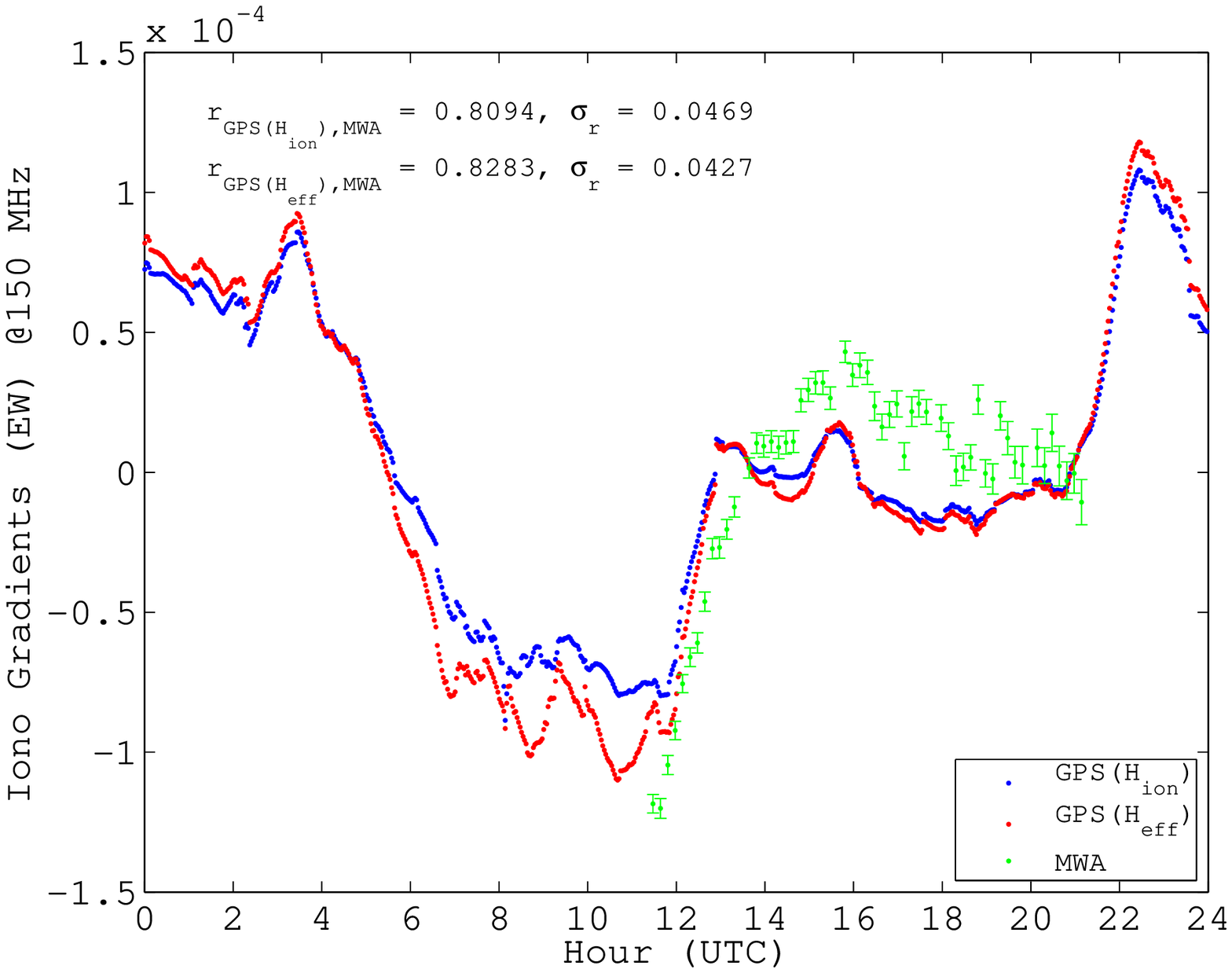}}\\ 
 \subcaptionbox{Single-station - GPS only, DOY 063\label{fig:EWMRO1063}}{\includegraphics[scale=0.33]{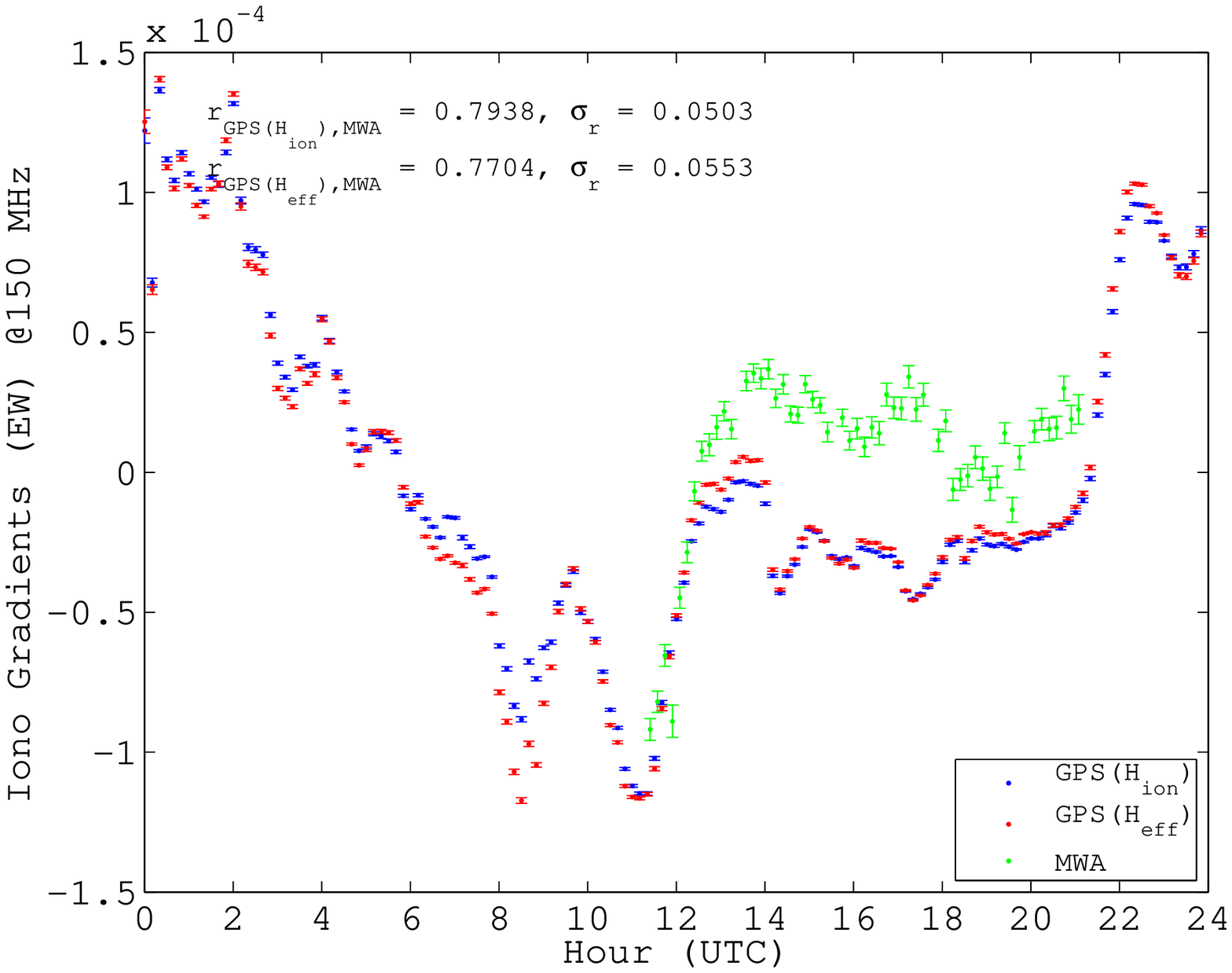}}
\subcaptionbox{Multi-station - GR, DOY 063 \label{fig:ewGR063}}{\includegraphics[scale=0.33]{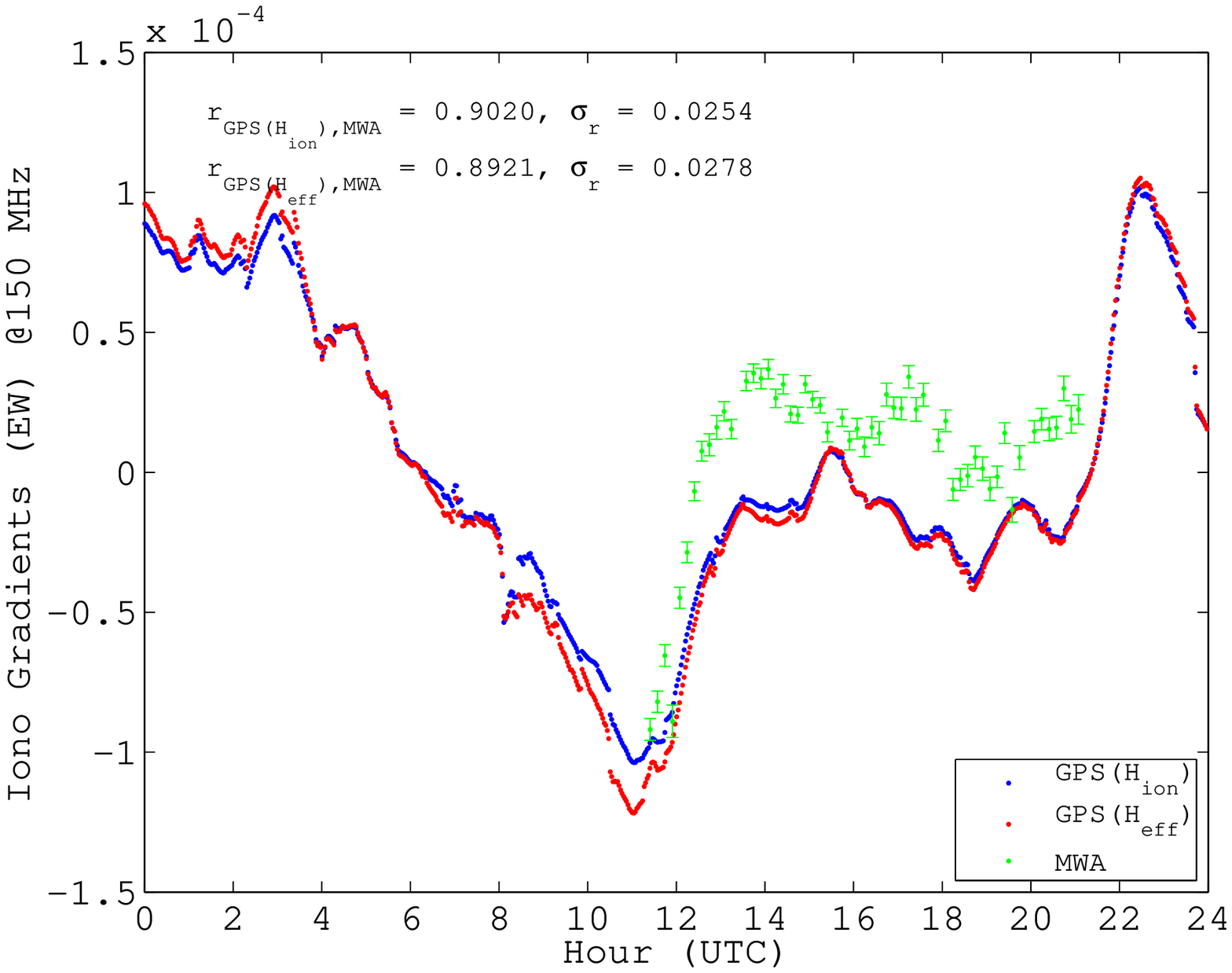}}\\ 
 \subcaptionbox{Single-station - GPS only, DOY 065\label{fig:EWMRO1065}}{\includegraphics[scale=0.33]{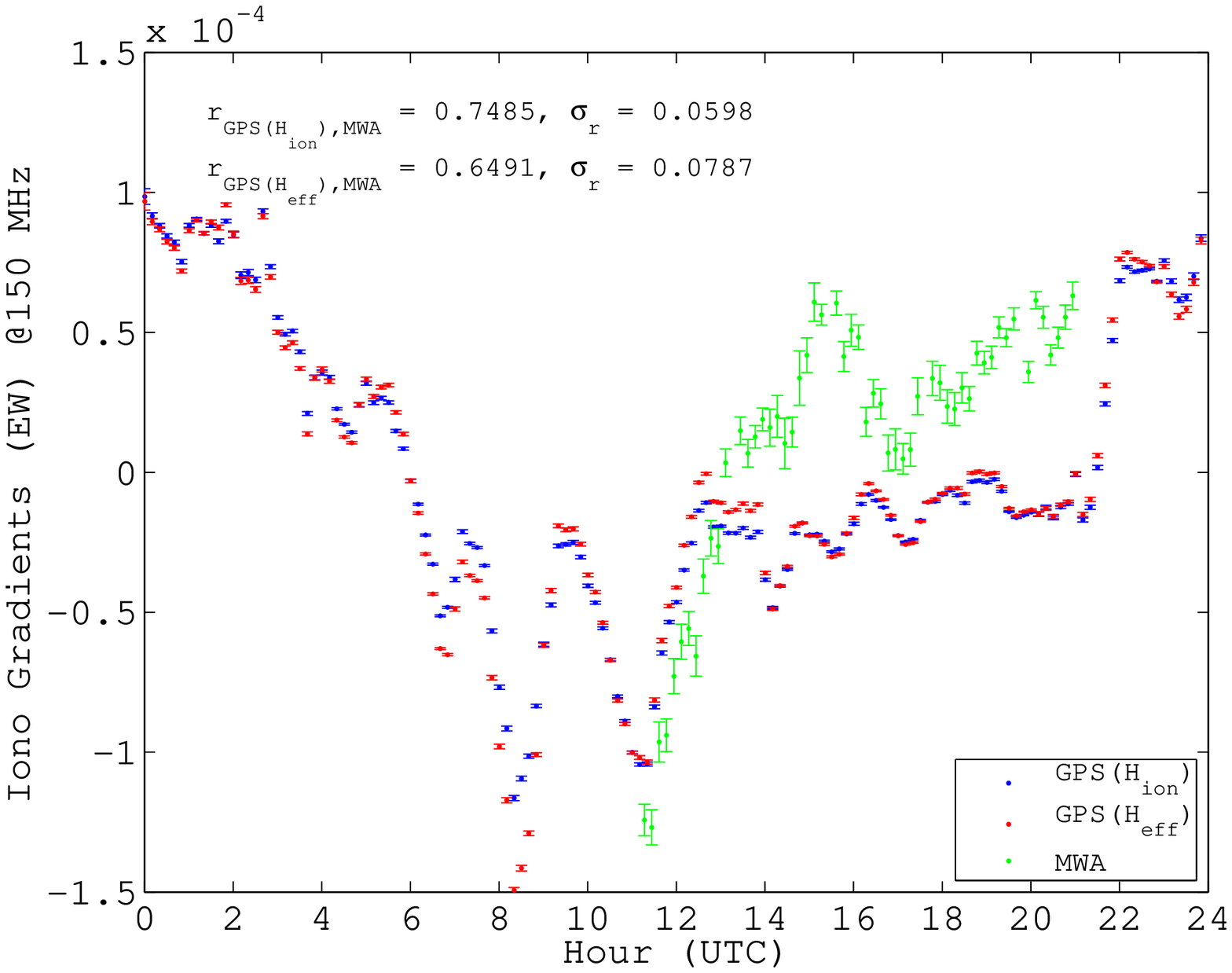}}
\subcaptionbox{Multi-station - GR, DOY 065 \label{fig:ewGR065}}{\includegraphics[scale=0.33]{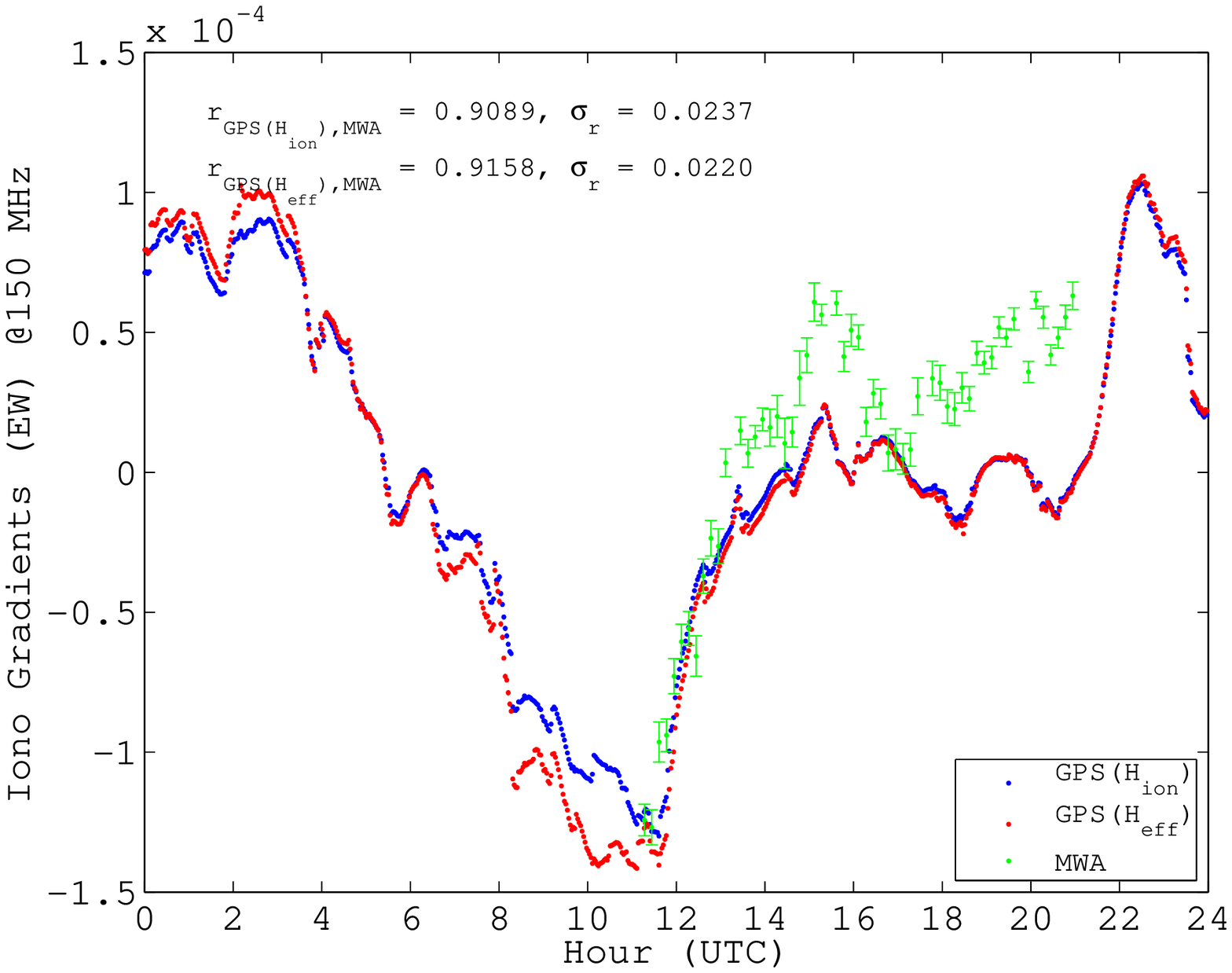}}\\ 
 \subcaptionbox{Single-station - GPS only, DOY 075\label{fig:EWMRO1075}}{\includegraphics[scale=0.33]{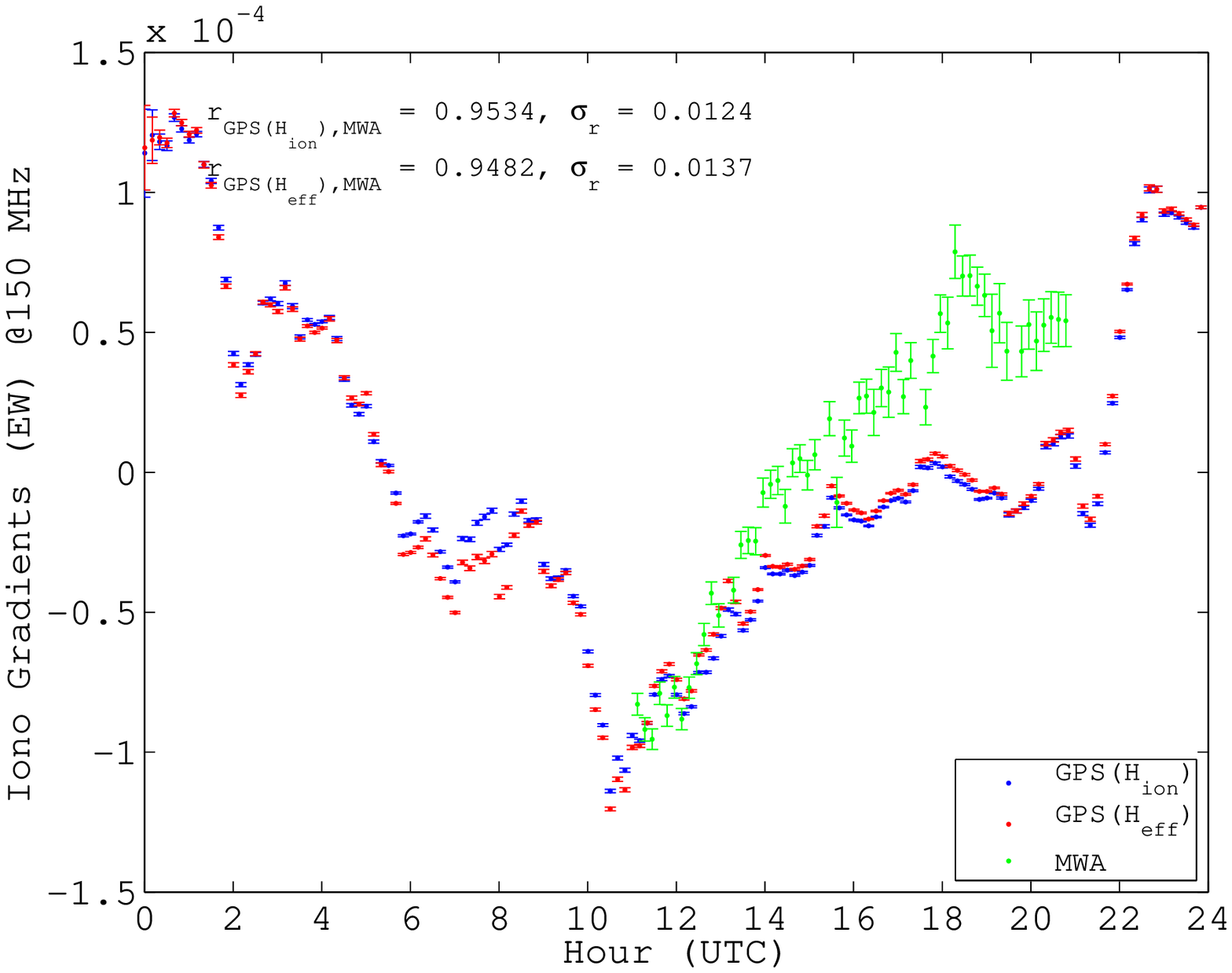}}
\subcaptionbox{Multi-station - GR, DOY 075 \label{fig:ewGR075}}{\includegraphics[scale=0.33]{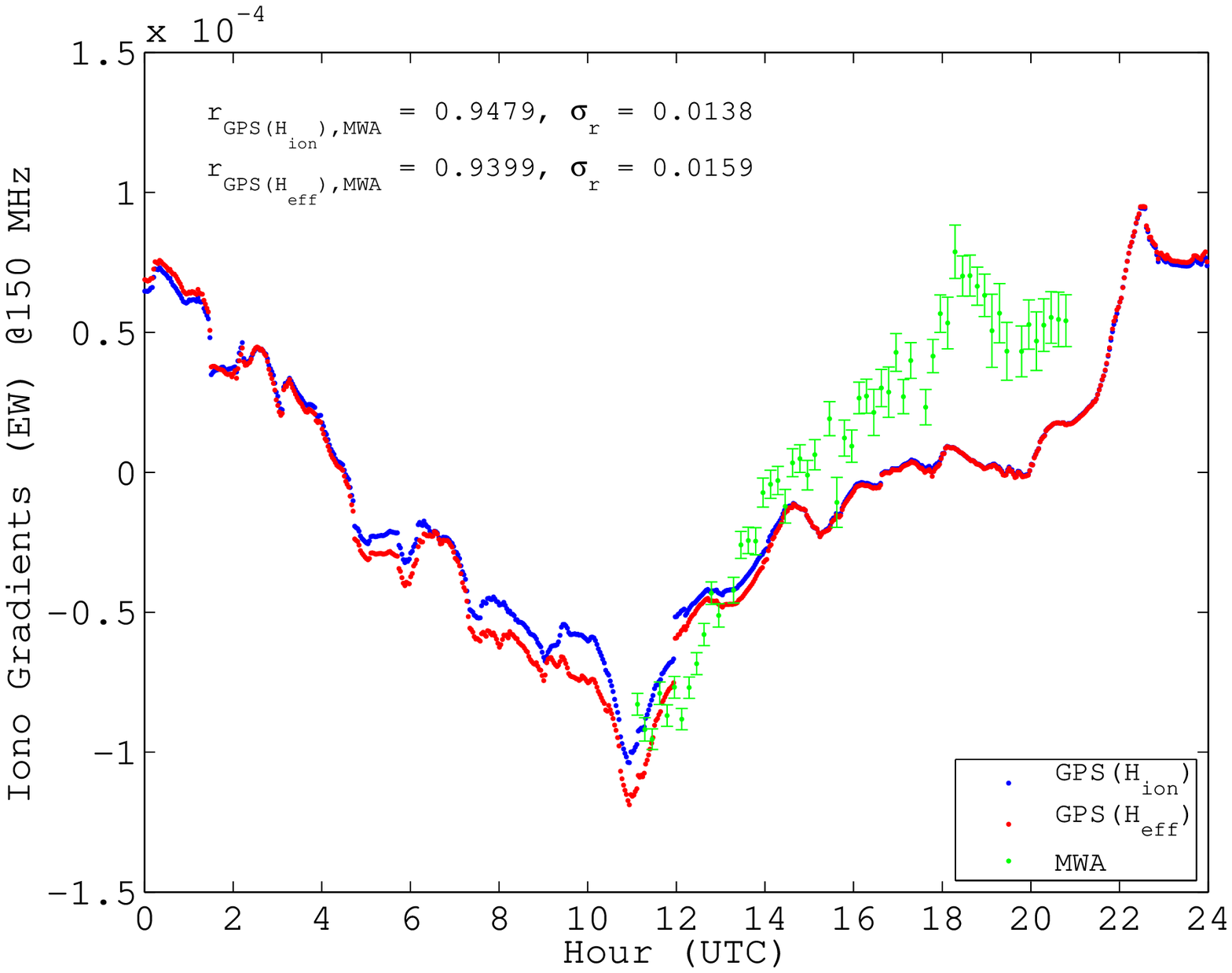}}
     \caption{EW ionosphere gradients observed from GNSS data [blue($H_{ion}$) and red($H_{eff}$)] and the MWA (green)  using single-station approach, GPS only (left column) and multi-station approach, GPS+GLONASS (GR, right column) on DOY 062, 063, 065 and 075, year 2014. Note the average precision of EW gradients is $\sim$0.07$\times 10^{-5}$ and $\sim$0.03$\times 10^{-5}$ for single-station and multi-station approach, respectively.
        \label{fig:offset2014ew}}
\end{figure*}

\begin{figure*}
 \centering
 \subcaptionbox{Single-station - GPS only, DOY 062\label{fig:NSMRO1062}}{\includegraphics[scale=0.33]{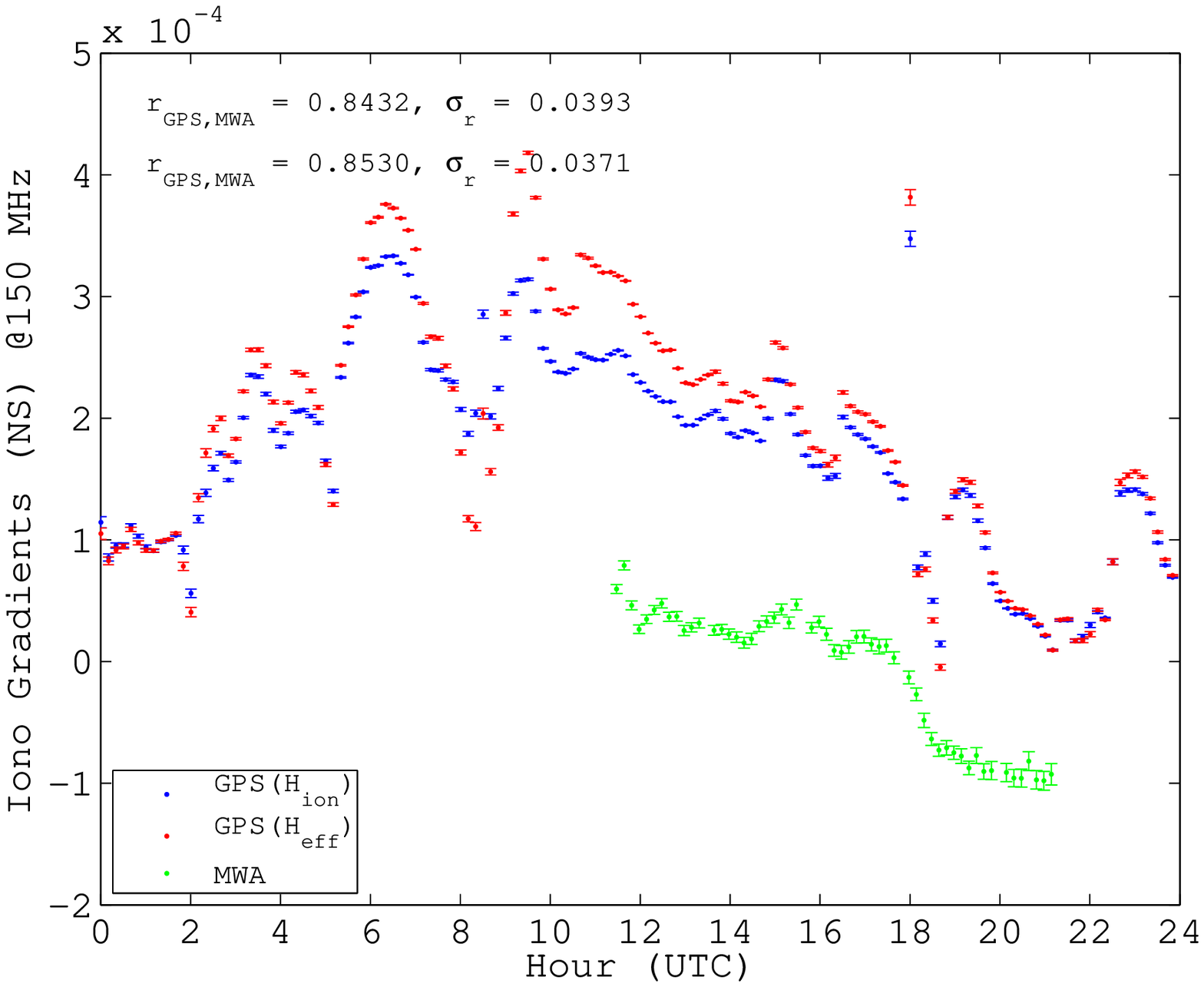}}
\subcaptionbox{Multi-station - GR, DOY 062 \label{fig:NSGR062}}{\includegraphics[scale=0.33]{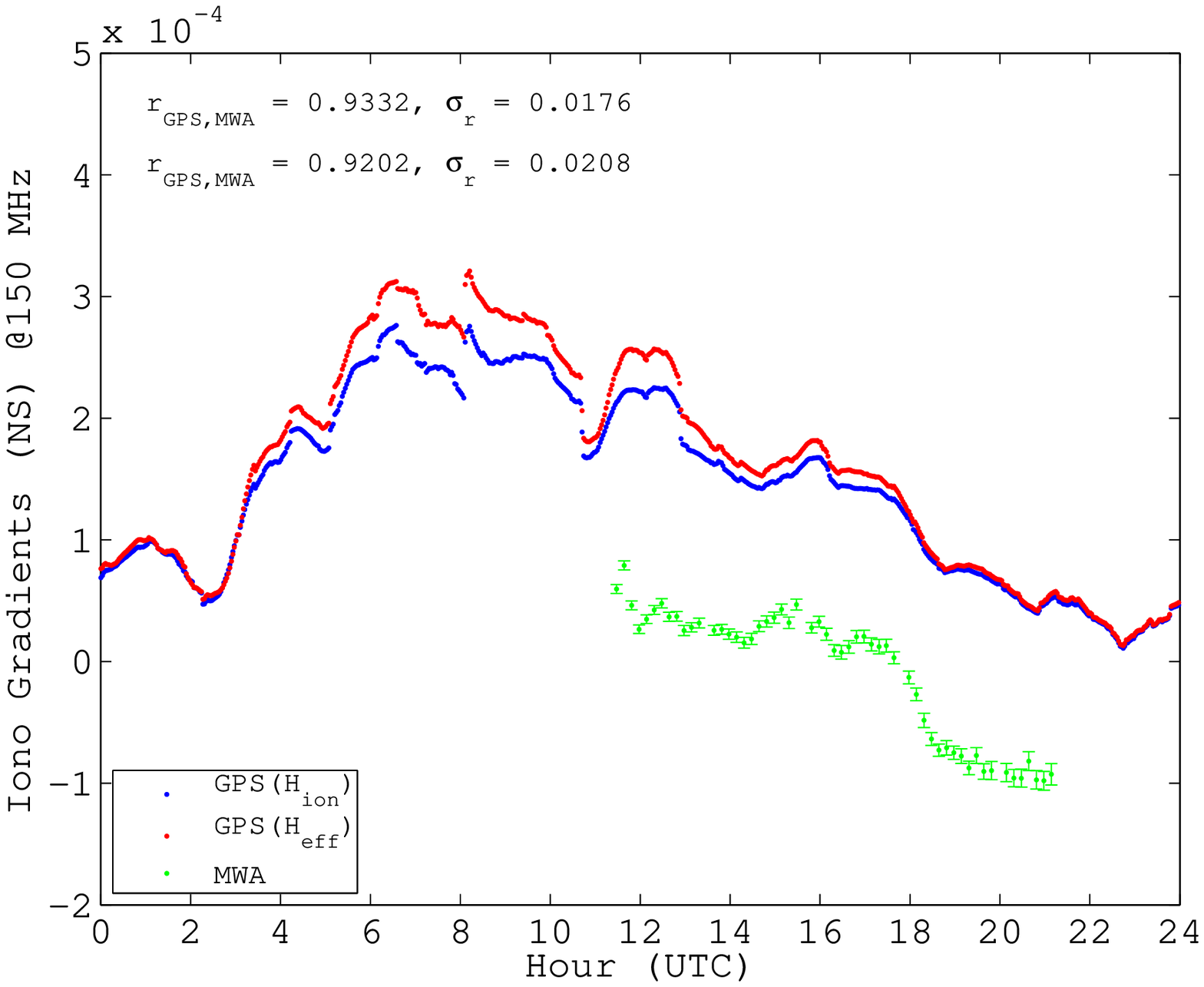}}
 \subcaptionbox{Single-station - GPS only, DOY 063\label{fig:NSMRO1063}}{\includegraphics[scale=0.33]{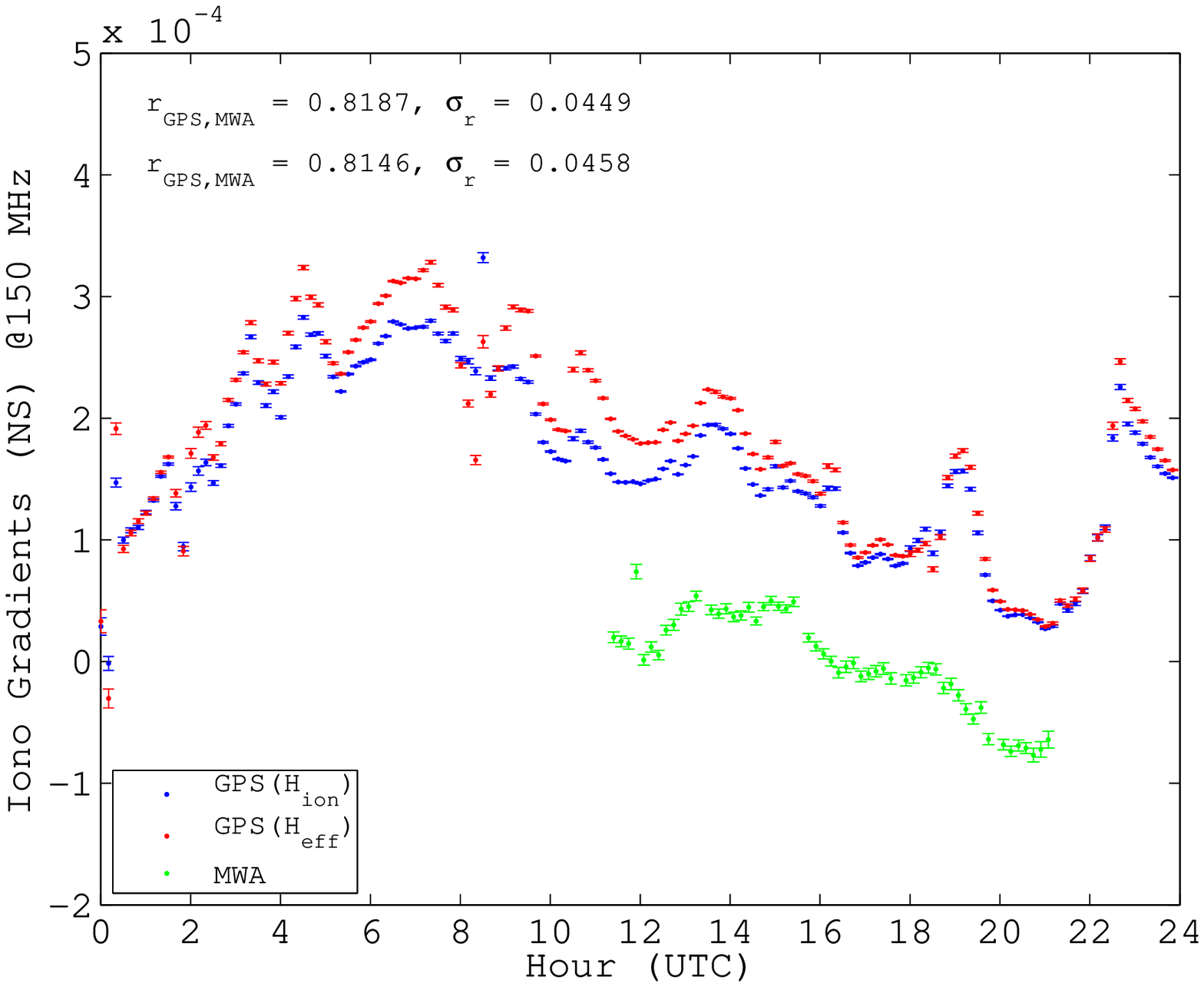}}
\subcaptionbox{Multi-station - GR, DOY 063 \label{fig:NSGR063}}{\includegraphics[scale=0.33]{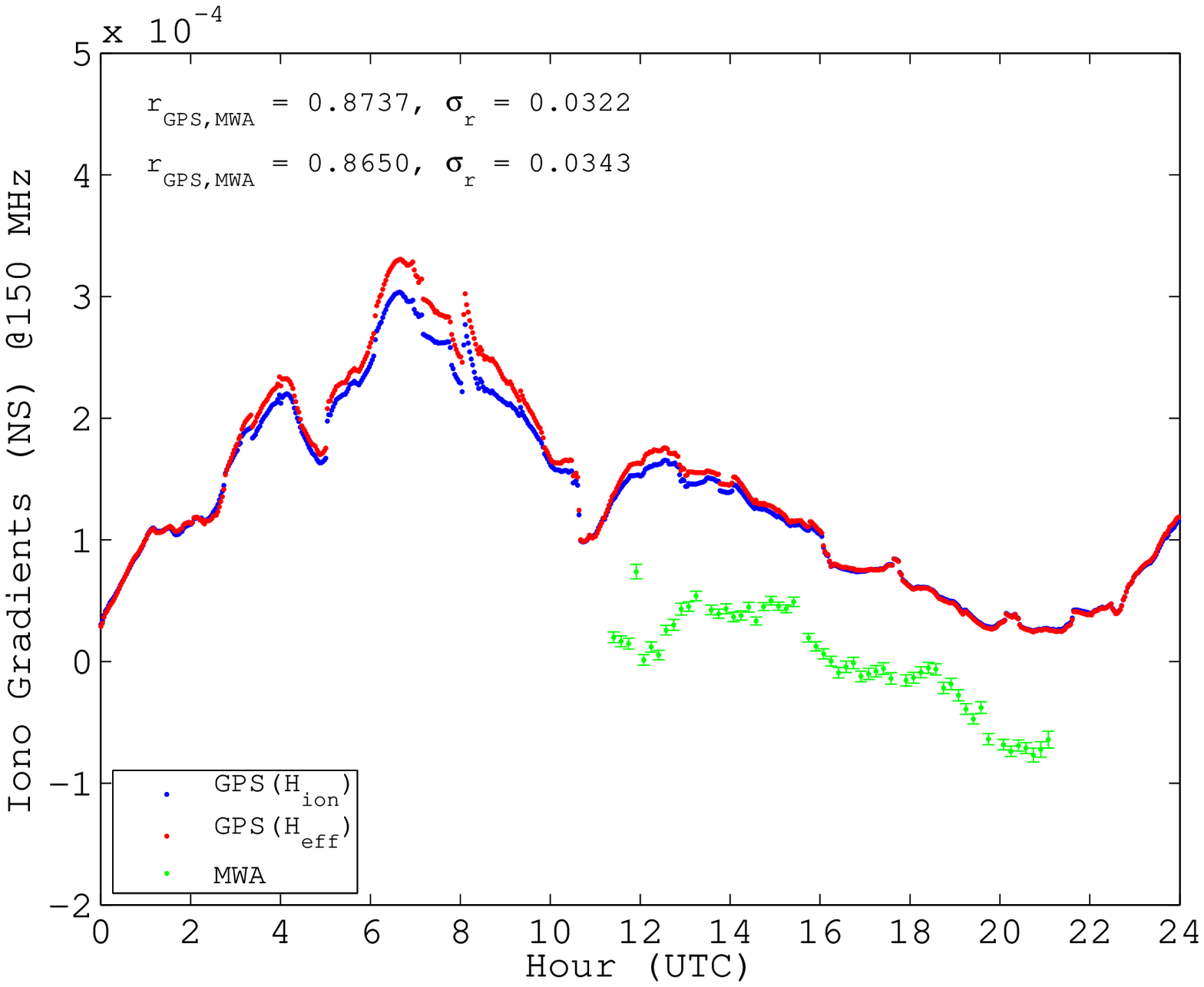}}
 \subcaptionbox{Single-station - GPS only, DOY 065\label{fig:NSMRO1065}}{\includegraphics[scale=0.33]{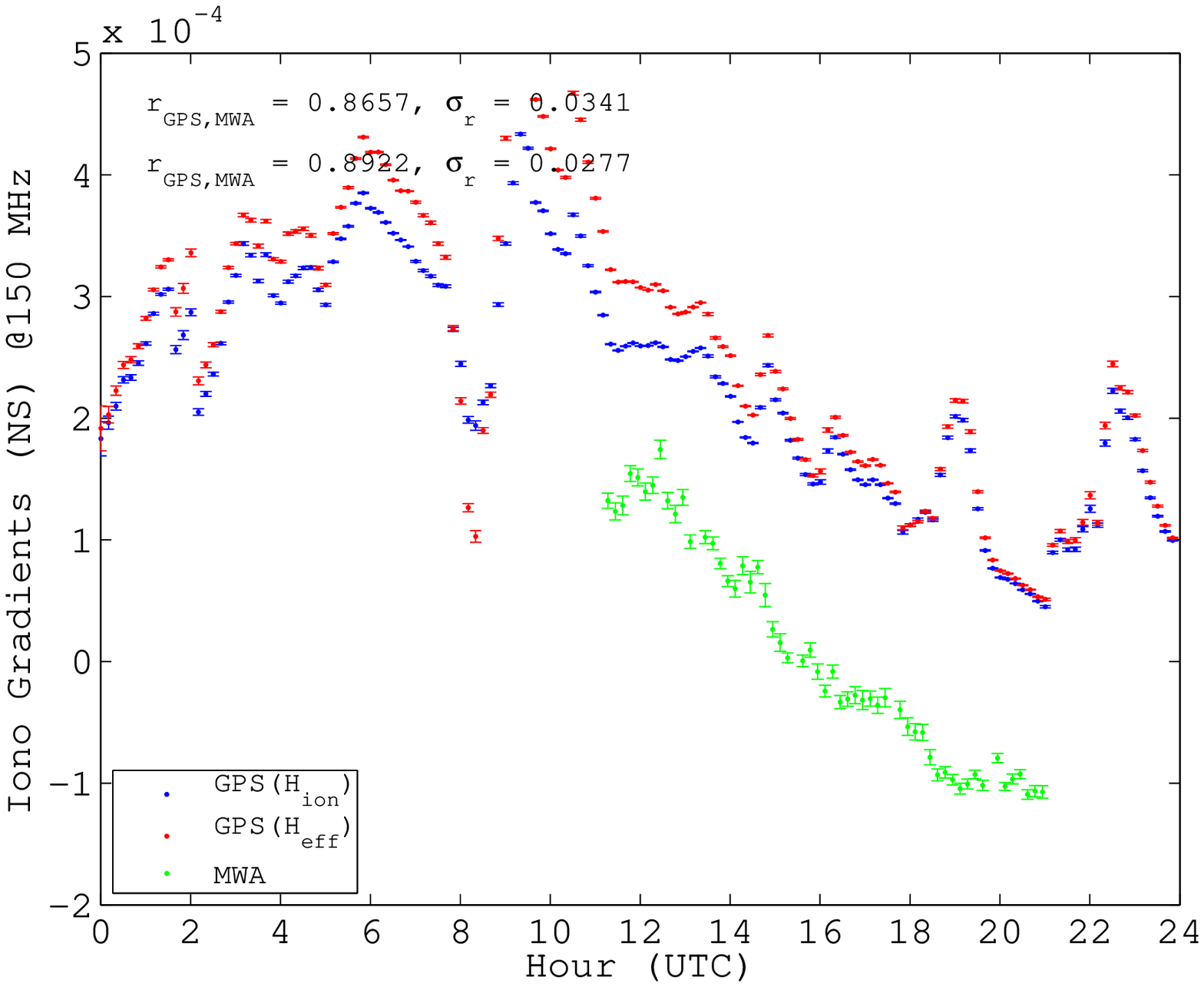}}
\subcaptionbox{Multi-station - GR, DOY 065 \label{fig:NSGR065}}{\includegraphics[scale=0.33]{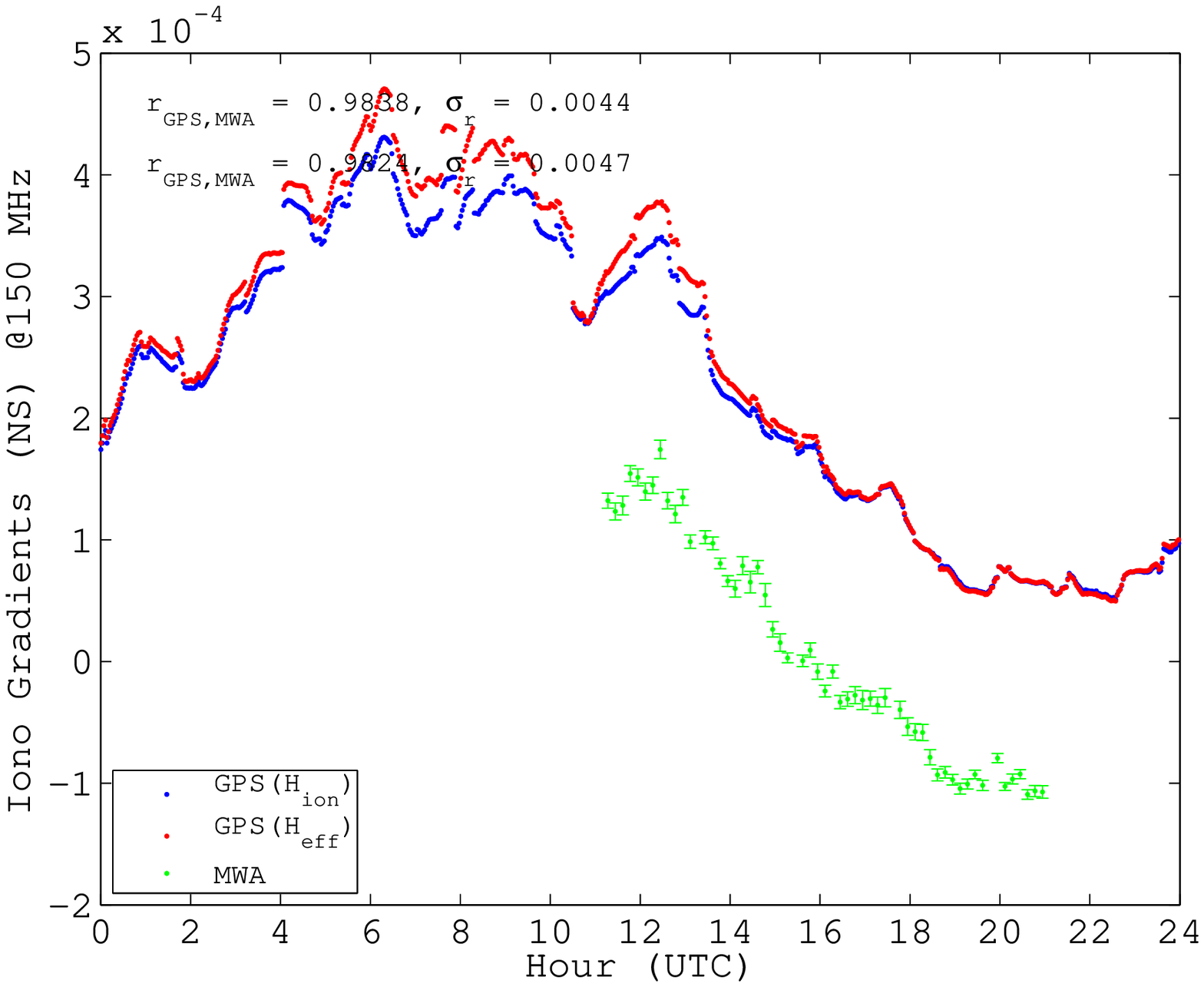}}
 \subcaptionbox{Single-station - GPS only, DOY 075\label{fig:NSMRO1075}}{\includegraphics[scale=0.33]{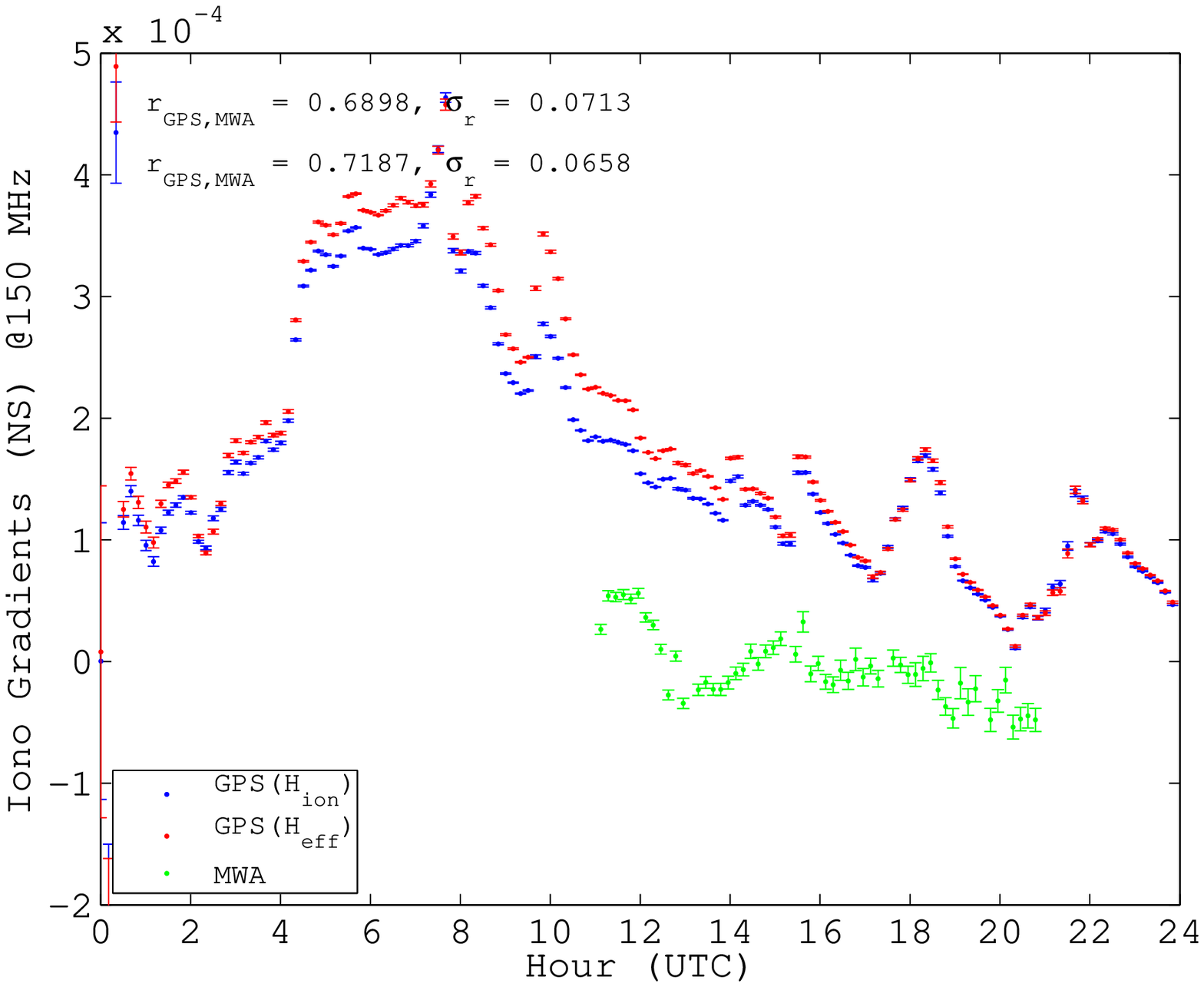}}
\subcaptionbox{Multi-station - GR, DOY 075 \label{fig:NSGR075}}{\includegraphics[scale=0.33]{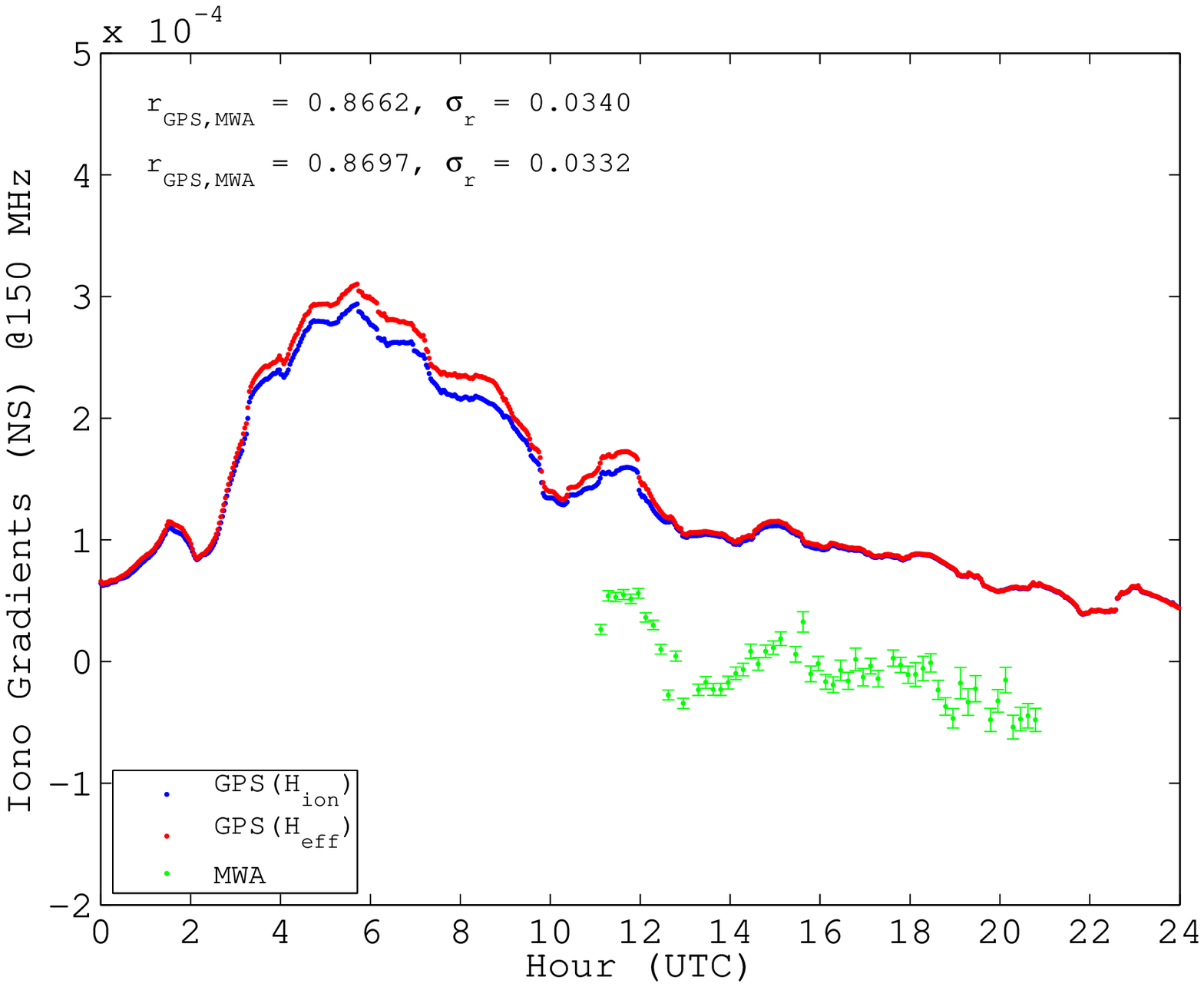}}
     \caption{NS ionosphere gradients observed from GNSS data (blue) and the MWA (green)  using single-station approach, GPS only (left column) and multi-station approach, GPS+GLONASS (GR, right column) on DOY 062, 063, 065 and 075, year 2014. Note the average precision of NS gradients is $\sim$0.05$\times 10^{-5}$ and $\sim$0.03$\times 10^{-5}$ for single-station and multi-station approach, respectively.
        \label{fig:offset2014ns}}
\end{figure*}

The MWA derived gradients are plotted (in green) along GNSS observed gradients for comparison. The ionospheric gradients, in EW and NS directions, were estimated for 03-03-2014 (DOY 062), 04-03-2014 (DOY 063), 06-03-2014 (DOY 065), and 16-03-2014 (DOY 075) using both single-station \citep{Aro15} and multi-station approaches (refer to Figures \ref{fig:offset2014ew} and \ref{fig:offset2014ns}). In each of the Figures \ref{fig:offset2014ew} and \ref{fig:offset2014ns}, two different ionosphere gradients are estimated, firstly considering a fixed height of the ionospheric layer ($H_{ion}$) at 450 km, plotted as blue line. Secondly, the height is assumed to vary in space and time, derived from IRI-Plas model ($H_{eff}$), indicated by red curve. \\

Table \ref{tab:gradcorr} presents the EW ($r_{EW}$) and NS ($r_{NS}$) gradient correlation between the GNSS and MWA observed gradients for the single-station and multi-station approach for all the days of observations.\\

The correlation between the GPS and MWA EW gradients are identical within the errors between the single-station and multi-station approaches for two of the days, refer Table \ref{tab:gradcorr}. For the remaining two days the EW gradient correlation was found to be significantly better using the multi-station approach. The NS gradient correlation was found to be significantly better for three of the four days using the multi-station approach. The general trend showed that the EW and NS gradients show better correlations with the MWA observed gradients when estimated using a multi-station approach rather than single-station approach. \\

The single-station approach is limited by the number of observations to constrain the gradients, hence the gradients appear to be noisy. This is however not the case with the multi-station approach. By using the model values for ionospheric shell height ($H_{eff}$) varying in space and time, a curvature is defined for the ionospheric layer, hence the gradients appear to have a steeper slope (Figures \ref{fig:offset2014ew} and \ref{fig:offset2014ns}), as compared to fixed ionospheric shell height ($H_{ion}$).

\begin{table*}
\caption{Correlation between the GNSS and MWA observed gradients in EW ($r_{EW}$) and NS ($r_{NS}$) components, its standard error ($\sigma_{r}$) using single-station approach and multi-station approach.} 
\begin{center}
\begin{tabular}{@{}c c c| c c c@{}}
\hline \hline
 DOY & \multicolumn{4}{c}{$ r_{EW}$ ($\sigma_{r_{EW}}$)}\\
     & \multicolumn{2}{c}{Single-station approach} & \multicolumn{2}{c}{Multi-station approach} \\ \cline{2-5}
      & $H_{ion}$ & $H_{eff}$ & $H_{ion}$ & $H_{eff}$ \\ \cline{2-5}
062  &  0.78(0.05) & 0.76(0.06) & 0.81(0.05) & 0.83(0.04)  \\
     &  &  &   &  &\\
063  &  0.79(0.05) & 0.77(0.05) & 0.90(0.03) & 0.89(0.03)  \\
     &  &  &  &  &\\
065  &  0.75(0.06) & 0.65(0.08) & 0.91(0.02) & 0.92(0.02)  \\
     &   &  &  &  &\\ 
075  &  0.95(0.01) & 0.95(0.01) & 0.95(0.01) & 0.94(0.02)  \\
     &  &  &  &  &\\ \hline
 	 & \multicolumn{4}{c}{$ r_{NS}$ ($\sigma_{r_{NS}}$)} \\ 
 	 & \multicolumn{2}{c}{Single-station approach} & \multicolumn{2}{c}{Multi-station approach} \\ \cline{2-5}
      & $H_{ion}$ & $H_{eff}$ & $H_{ion}$ & $H_{eff}$ \\ \cline{2-5}
062 & 0.84(0.04) & 0.85(0.04) & 0.93(0.02) & 0.92(0.02)  \\
     &  &  &   &  &\\
063 &  0.82(0.05) & 0.81(0.05) & 0.87(0.03) & 0.87(0.03)  \\
     &  &  &  &  &\\
065 & 0.87(0.03) & 0.89(0.03) & 0.98(0.01) & 0.98(0.01)  \\
     &   &  &  &  &\\ 
075 &  0.67(0.07) & 0.72(0.07) & 0.87(0.03) & 0.87(0.03)  \\
     &  &  &  &  &\\
\hline\hline
\end{tabular}
\end{center}
\label{tab:gradcorr}
\end{table*}

\section{CONCLUSIONS}
\label{sec:conc}

In this work, we have explored several incremental improvements on our previous work \citep{Aro15}, including: 1) the addition of GLONASS data to augment GPS data; 2) the development of a multi-station ionospheric solution rather than a single-station solution; and 3) the sensitivity of our analysis to varying height for a single layer ionospheric model. This work is designed to explore the most effective future directions for the development of ionospheric modelling to support calibration of the MWA and other future instruments.\\

The height of the single layer model is seen to play a significant role in the estimated ionosphere coefficients. The estimated ionosphere coefficients and receiver DCBs were seen to have a common minimum offset while the height of the single layer ($H_{ion}$) was varied. While a variable height of the single layer was incorporated using IRI-Plas model, the gradients appear to have a steeper slope. The increased steepness in the slope of the gradients could be due to curvature incorporated in the height of the single layer model ($H_{eff}$) by considering spatial and temporal variation.\\

For a single layer model, the height of the ionospheric layer is therefore an important parameter which influences the estimated coefficients. Also, the modelled $VTEC$ is limited by the obliquity factor used to map the $STEC$ to $VTEC$. Since the effective height of the ionosphere is known to vary both temporally and spatially, it is important to model the ionosphere using a three-dimensional spatial model.  We conclude that the future work should focus on the construction of a three-dimensional (or multi-layer) model for the ionosphere.\\

We have also found that the addition of GLONASS data to GPS data, and the use of a multi-station solution rather than a single-station solution, gives better results than our original work. The gradients from the multi-station approach were estimated at a higher time resolution (2 minutes) in comparison to single-station approach. Also, due to the large number of observations used to estimate gradients in a multi-station approach, the gradients seem to have a smoother temporal variation. For all the selected days of MWA observations, the correlation between MWA and GNSS estimated gradients was found to be identical within errors or higher with multi-station approach as compared to single-station approach.\\

The ionospheric modelling performed using GPS and GLONASS observations was also able to capture the spatial variations of the gradients, refer Figure \ref{fig:gradsepvar}. This encourages us towards deriving the higher order effects of the gradients estimated using GNSS. Future work will focus on deriving higher order effects in the gradients for various sources within MWA FoV and GNSS observations.\\

The NS gradients, estimated using the multi-station approach, agreed with the MWA observed gradients. The EW gradients had a better correlation than single-station approach, however did not seem to correlate as well as the NS gradients. The current distribution of GNSS receiving stations, while demonstrated to be successful in characterising large-scale ionospheric features and validating our technical approach, is inadequate for advanced modelling.\\
 
The MWA with its wide FoV imaging capability, sees a position offset due to the ionosphere for each of the sources in its FoV. This capability of the MWA has been exploited to detect small-scale structures in the ionosphere \citet{Loi2015a,Loi2015b,Loi2015c}. The spatial scales of the structures are around 10-100 km. These are precisely the spatial scales that will need to be characterised if future longer baseline instruments are to be calibrated.\\

GPS satellites are capable of providing ionospheric information, however the density of the pierce points is far too low to probe these small scales for the datasets currently available for the MRO. For a GPS receiver near the MRO, only 5-8 satellites are visible above horizon at any given time.  The number of measurements can be increased to 10-15 satellites by including data from GLONASS satellites. Future Global Navigation Satellite Systems (GNSS), namely, BeiDou (China) and Galileo (European Union) are expected to be operational around the year 2020 \citep{UN2010}. Regional satellite systems such as the Quasi-Zenith Satellite System (QZSS, Japan) \citep{UN2010}, currently under development, will have 4 satellites, all of which pass over the MRO. In this scenario 20-30 satellites would be above the horizon at the MRO at any given time, with an average separation of 200\,km on the ionosphere.\\

We have established a methodology to obtain ionospheric information with the current GNSS infrastructure around the MRO. This methodology could be exploited to derive ionospheric corrections for the future low frequency arrays, like the SKA-low, where the direction-dependent effects become more dominant and deriving ionospheric information from the astronomical observations may not be feasible. The GNSS-derived ionospheric information can also be used for climatology. This could be useful in designing future instruments, devising calibration strategies, and for selecting data post-observation (to avoid wasting effort on data which is too badly corrupted by the ionosphere). The current GNSS infrastructure which limits the spatial resolution of the ionospheric corrections can be improved by deploying additional GNSS receivers.\\

To measure the ionosphere on scales of $<$10 km, a dense cluster of GNSS receivers (5-10 receivers), with baselines as small as $\sim$5 km would need to be installed. This would allow the ionospheric gradients, rather than just the $STEC$, to be determined towards each satellite. However, it is not guaranteed that a GNSS satellite would be in the MWA field of view at all times. Hence another small cluster of GNSS stations would need to be deployed strategically. Deploying the further cluster at a distance of $\sim$100 km would fill in the gaps between existing satellites. We call this approach the ``cluster of clusters'' approach. For further densification, a cluster of GNSS stations at the median of existing clusters ($\sim$50 km) could be deployed.\\

\begin{figure}[h]
 \centering
{ \includegraphics[scale=0.6]{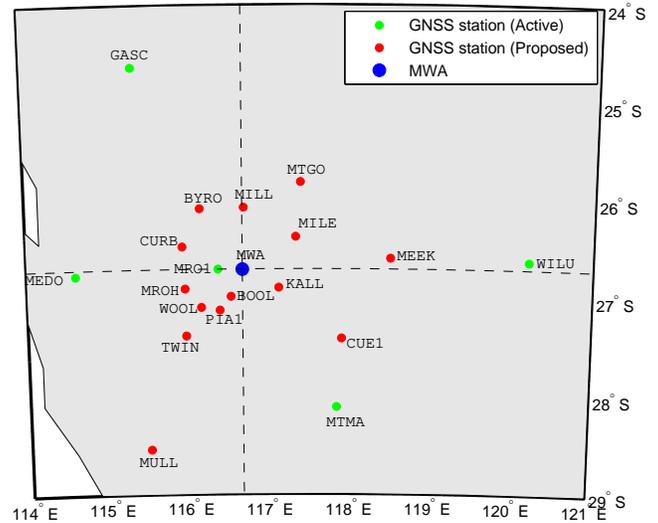}}  
     \caption{Current (green) and proposed (red) GNSS station locations in vicinity of MRO. The MWA location is marked in blue.
        \label{fig:propGPSstn}}
\end{figure}

The sparse population around the MRO and the lack of remote power and communication infrastructure constrains the possible locations for such a cluster. Locations indicated in Figure \ref{fig:propGPSstn} are plausible cluster locations given that they are existing communities and homesteads likely to have the necessary infrastructure. \\

Future work will evaluate the expansion of GNSS stations around MRO, in view of generating a three-dimensional ionospheric model, to meet the ionosphere calibration requirements for future MRO instruments.\\ 

\section{ACKNOWLEDGEMENTS}
\label{sec:ack}
The authors wish to thank John Kennewell from Curtin University for the valuable discussions. This scientific work makes use of the Murchison Radio-astronomy Observatory, operated by CSIRO. We acknowledge the Wajarri Yamatji people as the traditional owners of the Observatory site. Support for the operation of the MWA is provided by the Australian Government Department of Industry and Science and Department of Education (National Collaborative Research Infrastructure Strategy: NCRIS), under a contract to Curtin University administered by Astronomy Australia Limited. We acknowledge the iVEC Petabyte Data Store and the Initiative in Innovative Computing and the CUDA Center for Excellence sponsored by NVIDIA at Harvard University.

\newcommand{\pasa}{PASA}
\newcommand{\aj}{AJ}
\newcommand{\apj}{ApJ}
\newcommand{\apjs}{ApJS}
\newcommand{\apjl}{ApJL}
\newcommand{\aap}{A{\&}A}
\newcommand{\aaps}{A{\&}AS}
\newcommand{\mnras}{MNRAS}
\newcommand{\araa}{ARAA}
\newcommand{\pasp}{PASP}
\bibliographystyle{apj}
\bibliography{./Arora_et_al_2016} 

\end{document}